\documentclass[%
 aip,
 amsmath,amssymb,
 reprint,%
]{revtex4-1}

\usepackage{graphicx}
\usepackage{dcolumn}
\usepackage{bm}
\usepackage{orcidlink}
\usepackage[utf8]{inputenc}
\usepackage[T1]{fontenc}
\usepackage{mathptmx}
\usepackage{etoolbox}
\usepackage{color}
\usepackage{hyperref}
\usepackage{float}
\usepackage{algorithm}
\usepackage{algpseudocode}
\algrenewcommand\algorithmicrequire{\textbf{Input:}}
\algrenewcommand\algorithmicensure{\textbf{Output:}}
\usepackage{plimsoll}
\usepackage{siunitx}
\usepackage{CJKutf8}
\usepackage{nomencl}
\makenomenclature
\renewcommand\nomgroup[1]{%
  \item[\bfseries
  \ifstrequal{#1}{V}{Symbols}{%
  \ifstrequal{#1}{S}{Subscripts}{%
  \ifstrequal{#1}{A}{Acronyms}{}}}%
]}

\makeatletter
\def\@email#1#2{%
 \endgroup
 \patchcmd{\titleblock@produce}
  {\frontmatter@RRAPformat}
  {\frontmatter@RRAPformat{\produce@RRAP{*#1\href{mailto:#2}{#2}}}\frontmatter@RRAPformat}
  {}{}
}%
\makeatother
\begin{document}

\preprint{AIP/123-QED}

\title[]{A Unified Pore-Scale Multiphysics Model for the Integrated Soot Transport-Deposition-Oxidation in Catalytic Diesel Particulate Filters}

\author{Yujing Zhang~(\begin{CJK*}{UTF8}{gbsn}张宇靖\end{CJK*})~\orcidlink{0009-0002-8800-1341}}
\affiliation{School of Automotive Studies, Tongji University, Shanghai 201804, China.}

\author{Yunhua Zhang~(\begin{CJK*}{UTF8}{gbsn}张允华\end{CJK*})~\orcidlink{0000-0003-1968-9365}}
\email[Corresponding email:~]{zhangyunhua@tongji.edu.cn}
\affiliation{School of Automotive Studies, Tongji University, Shanghai 201804, China.}

\author{Liang Fang~(\begin{CJK*}{UTF8}{gbsn}房亮\end{CJK*})~\orcidlink{0000-0002-3943-3175}}
\affiliation{School of Automotive Studies, Tongji University, Shanghai 201804, China.}

\author{Diming Lou~(\begin{CJK*}{UTF8}{gbsn}楼狄明\end{CJK*})}
\affiliation{School of Automotive Studies, Tongji University, Shanghai 201804, China.}

\author{Piqiang Tan~(\begin{CJK*}{UTF8}{gbsn}谭丕强\end{CJK*})}
\affiliation{School of Automotive Studies, Tongji University, Shanghai 201804, China.}

\author{Zhiyuan Hu~(\begin{CJK*}{UTF8}{gbsn}胡志远\end{CJK*})}
\affiliation{School of Automotive Studies, Tongji University, Shanghai 201804, China.}

\date{\today}

\begin{abstract}
    Understanding the intricate interplay between soot dynamics and chemical reactions within catalytic diesel particulate filters (CDPF) is crucial for enhancing both filtration efficiency and regeneration performance. In this paper, we establish a unified pore-scale multiphysics model based on the Eulerian-Lagrangian framework to comprehensively resolve the transport, deposition, and oxidation of soot. Distinguishing itself from conventional empirical correlations and stochastic-based approximations, the system models soot deposition through fundamental physical principles, integrating elastic deformation and surface adhesion mechanics at the particle-wall interface. Simultaneously, it incorporates a robust oxidation model that accounts for the competitive kinetics of both $\textrm{O}_2$ and $\textrm{NO}_2$ pathways, enabling comprehensive coverage of all CDPF operating regimes. Validated against three classical benchmark cases, the model demonstrates superior accuracy in capturing interfacial mass transfer and particle-wall interactions. Simulation under a typical CDPF low-temperature operating condition emphasizes the pivotal role of $\textrm{NO}_2$ and catalyst in promoting regeneration and reveals complex synergistic and competitive effects between distinct reaction pathways. Notably, the reaction rate of direct $\textrm{O}_2$ pathway is accelerated by a factor of 87 in the presence of the catalyst. For ultra-fine soot particles ($50~\mathrm{nm}$), the Brownian motion and thermophoretic forces directly dictate the deposition efficiency. Their strong thermal sensitivity also underscores the necessity for an integrated soot transport-deposition-oxidation framework. To support further research, the model implementation can be accessed at \href{https://github.com/zhangyujing2001}{GitHub}.
\end{abstract}

\maketitle


\nomenclature[V]{$u$}{Velocity,~$\mathrm{m/s}$}
\nomenclature[V]{$t$}{Time,~$\mathrm{s}$}
\nomenclature[V]{$p$}{Pressure,~$\mathrm{Pa}$}
\nomenclature[V]{$g$}{Gravitational acceleration,~$\mathrm{m/s^{2}}$}
\nomenclature[V]{$Y$}{Mass fraction,~$-$}
\nomenclature[V]{$D$}{Mass diffusivity,~$\mathrm{m^2/s}$}
\nomenclature[V]{$\dot m$}{Mass production rate,~$\mathrm{kg/(m^{i}\cdot s)}$,~$\mathrm{i=2}$ for heterogeneous reaction,~$\mathrm{i=3}$ for homogeneous reaction}
\nomenclature[V]{$M$}{Molar mass,~$\mathrm{kg/mol}$}
\nomenclature[V]{$v_\textrm{r}$}{Reaction rate,~$\mathrm{mol/(m^{i}\cdot s)}$,~$\mathrm{i=2}$ for heterogeneous reaction,~$\mathrm{i=3}$ for homogeneous reaction}
\nomenclature[V]{$T$}{Temperature,~$\mathrm{K}$}
\nomenclature[V]{$C_p$}{Specific heat capacity,~$\mathrm{J/(kg\cdot K)}$}
\nomenclature[V]{$k$}{Thermal conductivity,~$\mathrm{W/(m\cdot K)}$}
\nomenclature[V]{$q^{\prime \prime}$}{Heat flux,~$\mathrm{W/m^{2}}$}
\nomenclature[V]{$\Delta_\textrm{r}H^{\plimsoll}$}{Standard reaction enthalpy change,~$\mathrm{J/mol}$}
\nomenclature[V]{$\dot q_V$}{Rate of heat production, $\mathrm{W/m^3}$}
\nomenclature[V]{$A_\textrm{h}$}{Hamaker constant,~$-$}
\nomenclature[V]{$Z_\textrm{sep}$}{Minimum separation distance,~$0.4~\mathrm{nm}$}
\nomenclature[V]{$Z_\textrm{sep,0}$}{Cut-off separation distance,~$0.165~\mathrm{nm}$}
\nomenclature[V]{$F$}{Force,~$\mathrm{N}$}
\nomenclature[V]{$e$}{Restitution coefficient,~$-$}
\nomenclature[V]{$c$}{Molar concentration,~$\mathrm{mol/m^3}$}
\nomenclature[V]{$Re$}{Reynolds number,~$-$}
\nomenclature[V]{$Da$}{Damköhler number,~$-$}
\nomenclature[V]{$Fo$}{Fourier number,~$-$}
\nomenclature[V]{$Sh$}{Sherwood number,~$-$}
\nomenclature[V]{$Sc$}{Schmidt number,~$-$}
\nomenclature[V]{$St$}{Stokes number,~$-$}
\nomenclature[V]{$K$}{Permeability,~$\mathrm{m^2}$}
\nomenclature[V]{$\phi$}{Porosity,~$-$}
\nomenclature[V]{$H$}{Domain width,~$\mathrm{m}$}
\nomenclature[V]{$L$}{Domain length,~$\mathrm{m}$}
\nomenclature[V]{$A$}{Frequency factor,~$\mathrm{s^{-1}}$}
\nomenclature[V]{$E_a$}{Activation energy,~$\mathrm{J/mol}$}
\nomenclature[V]{$m$}{Mass,~$\mathrm{kg}$}
\nomenclature[V]{$d$}{Diameter,~$\mathrm{m}$}
\nomenclature[V]{$C_c$}{Cunningham correction factor,~$-$}
\nomenclature[V]{$R$}{Gas constant,~$\mathrm{J/(mol\cdot K)}$}
\nomenclature[V]{$Kn$}{Knudsen number,~$-$}
\nomenclature[V]{$K_\textrm{eq}$}{Thermodynamic equilibrium constant,~$-$}
\nomenclature[V]{$\alpha$}{thermal diffusivity,~$\mathrm{m^2/s}$; reaction order,~$-$}
\nomenclature[V]{$\rho$}{Density,~$\mathrm{kg/m^3}$}
\nomenclature[V]{$\mu$}{Dynamic viscosity,~$\mathrm{Pa\cdot s}$}
\nomenclature[V]{$\nu$}{Kinematic viscosity,~$\mathrm{m^2/s}$; stoichiometric number,~$-$}
\nomenclature[V]{$\eta$}{Impaction efficiency,~$-$}
\nomenclature[V]{$\varphi$}{Physical quantity}
\nomenclature[V]{$\tau$}{Time scale,~$\mathrm{s}$}
\nomenclature[V]{$\gamma$}{Surface energy,~$\mathrm{N/m}$}

\nomenclature[S]{f}{fluid}
\nomenclature[S]{p}{Particle}
\nomenclature[S]{r}{Reaction}
\nomenclature[S]{eff}{Effective}
\nomenclature[S]{eq}{Equilibrium}
\nomenclature[S]{conv}{Convection}
\nomenclature[S]{diff}{Diffusion}
\nomenclature[S]{cr}{Critical}

\nomenclature[A]{PM}{Particulate matter}
\nomenclature[A]{CDPF}{Catalytic diesel particulate filter}
\nomenclature[A]{DPF}{Diesel particulate filter}
\nomenclature[A]{CFD}{Computational fluid dynamics}
\nomenclature[A]{XRT}{X-ray tomography}
\nomenclature[A]{FVM}{Finite volume method}
\nomenclature[A]{LBM}{Lattice Boltzmann method}
\nomenclature[A]{SEM}{Scanning electron microscope}
\nomenclature[A]{UDF}{User-defined function}
\nomenclature[A]{UDM}{User-defined memory}
\nomenclature[A]{UDS}{User-defined scalar}
\nomenclature[A]{GCI}{Grid convergence index}

\printnomenclature

\section{Introduction} \label{sec:introduction}
Due to high torque output \cite{wang2025comparative} and superior durability \cite{jathar2025short}, diesel engines still account for a significant proportion of power systems used in heavy-duty machinery \cite{huang2019effects}. However, particulate matter (PM) emissions have become a major constraint on their future development \cite{ansari2024experimental}. Compared with gasoline engines, the PM emissions from diesel engines are higher by a factor of 30 to 80 \cite{zhang2024particle}. Moreover, the lifespan of diesel exhaust PM in the atmosphere varies from minutes to days \cite{hime2018comparison}, which poses a substantial threat to human health \cite{fazakas2024health}. The latest Euro 7 emission standard \cite{dornoff2024euro, barbier2024analysis} reduces the minimum detectable PM diameter from $23~\mathrm{nm}$ to $10~\mathrm{nm}$ to better control ultra-fine nucleation mode particles in diesel exhaust \cite{zare2020emissions}, placing more stringent demands on diesel exhaust after-treatment technology \cite{recsitouglu2015pollutant}.

The catalytic diesel particulate filter (CDPF) \cite{van2001science}, a key component of diesel exhaust aftertreatment system \cite{ayodhya2018overview}, is regarded as one of the most promising approaches for PM control \cite{nakagoshi2023new} by integrating a diesel particulate filter (DPF) \cite{rodriguez2017regeneration} with catalysts to capture soot particles and facilitate its low-temperature oxidation \cite{zhang2024effect}. Accordingly, the filtration and regeneration performance become critical metrics in CDPF design \cite{li2025multi}.

Although conventional engine bench tests can dynamically assess the macroscopic filtration and regeneration performance of a CDPF by comparing particle emission characteristics at the inlet and outlet \cite{zhang2020study,zhang2023effect,bao2025experimental,lou2022analysis,rossomando2019experimental,hu2021experimental}, they cannot quantitatively elucidate the effect of different capture mechanisms that constitute the theoretical foundations of the entire filtration process \cite{guedes2009deep,zhang2022simplified,liu2020progress}. In addition, experiments on soot-catalyst oxidation, while useful for evaluating intrinsic reaction kinetics and catalyst activity \cite{huang2024study,zhang2024catalytic,gao2024effect,lv2025atomic,zhao2024advancements,liu2025acid}, are generally conducted under idealized conditions, which differ significantly from the complex physicochemical environment in practical CDPF operation \cite{kuang2024effect}.

With the advancement of computer hardware and numerical algorithms \cite{singh2025evolution,patawari2025traditional}, computational fluid dynamics (CFD) has become a powerful tool for analyzing complex transport phenomena in porous media systems \cite{simonov2023review}, such as the multiphase displacement dynamics in geoscience \cite{maes2022geochemfoam,singh2017dynamics,guiltinan2021two}, and the thermal protection technology in hypersonic aerodynamics \cite{zheng2025pore,zhang2025hybrid}. Nevertheless, compared with the successful applications in aforementioned areas, CFD-based studies of pore-scale CDPF are still at an early stage. Ko\v{c}\'{i} and his team employed time-resolved X-ray tomography (XRT) to reconstruct the three-dimensional porous structure of CDPFs with different catalyst loadings \cite{vaclavik2017structure,blavzek2021washcoating}, and on this basis investigated the flow behavior inside porous media \cite{kovci20193d,leskovjan2021multiscale} using OpenFOAM, an open-source CFD code based on the finite volume method (FVM) \cite{jasak2007openfoam}. To further explore the effect of soot layer on the reaction-diffusion process, N\v{e}mec and Ko\v{c}\'{i} \cite{nemec2025effective} established an effective one-dimensional mathematical model based on a classical DPF wall model. The simulation results from this model suggest that soot layer has no effect on reactions. Prior to the Ko\v{c}\'{i}'s team, research group led by Konstandopoulos \cite{konstandopoulos2000fundamental,vlachos2006digital,konstandopoulos2012aspects} first simulated the fluid flow inside a CDPF generated by the fractional Brownian motion method using the lattice Boltzmann method (LBM). Following this approach, Yamamoto and his colleagues \cite{yamamoto2011lattice,yamamoto2013numerical,yamamoto2015simulation} simulated the soot oxidation process in a CDPF under high-temperature conditions, in which only one reaction was considered. In recent years, increasing attention has also been paid to the deposition behavior of soot particles within CDPFs. Existing models are typically based either on the Eulerian approach, which treats the particle phase as a continuum and derives the filtration efficiency from concentration variations \cite{yamamoto2015simulation,kong2019simulation}, or on cellular automata (CA) probabilistic model, which adopts a snowfall-like analogy \cite{chopard1999cellular} to simulate the deposition process \cite{kong2019numerical,huang2025simulation}. Despite these advances, several fundamental challenges persist:

(1)~Existing deposition models oversimplify particle-wall interactions, relying either on empirical correlations from bench tests \cite{kong2019simulation,tan2019modeling,li2023study} or stochastic frameworks that lack mechanistic depth \cite{kong2019numerical,belot2020numerical,huang2024study}.

(2)~While pore-scale hydrodynamics has been extensively explored, relatively little attention has been paid to the coupled heat and mass transfer phenomena induced by chemical reactions, especially for low-temperature soot oxidation process.

(3)~Previous studies typically treat deposition and oxidation as decoupled processes, which contradicts the actual concurrent operating scenarios of CDPF systems.

Considering the limitations of previous studies, in this work, we develop a unified pore-scale multiphysics model based on the Eulerian-Lagrangian framework \cite{patankar2001modeling} to simulate the complex transport phenomena of CDPF. Three contributions are presented in this paper. First, a readily implementable soot oxidation model is derived based on chemical kinetics and thermochemistry, which fully accounts for CDPF regeneration behavior under low-temperature operating conditions. Second, based on the elastic deformation and surface adhesion theories, we establish a mathematical model to describe the interaction between soot particles and walls, which provides a physically grounded interpretation of the particle collision–adhesion–rebound process and demonstrates the influence of different impaction mechanisms during the filtration stage. Third, we focus on soot deposition under oxidative atmospheres, that is, the transport, deposition, and oxidation behaviors of soot are systematically integrated and investigated.

For the remaining sections, this paper is organized as follows: we first present the mathematical formulation of the pore-scale CDPF multiphysics coupling system in Sec.~\ref{sec:mathematical formulation}. Next, we conduct several benchmark-based verifications on our code to confirm its capability of capturing complex interfacial transfer phenomena and the particle-wall interactions in the established model (Sec.~\ref{sec:benchmark based code verification}). Then, the proposed model is applied to simulate the regeneration and particulate filtration processes in an in-wall CDPF under a typical operating condition and the results are discussed in Sec.~\ref{sec:results and discussion}. Finally, conclusions of our current work and future perspectives are presented in Sec.~\ref{sec:conclusion}.

\section{Mathematical Formulation} \label{sec:mathematical formulation}

\subsection{Geometry representation of the CDPF porous structure} \label{sec:cdpf geom}
\begin{figure}[ht]
    \centering
    \includegraphics[width=0.65\linewidth]{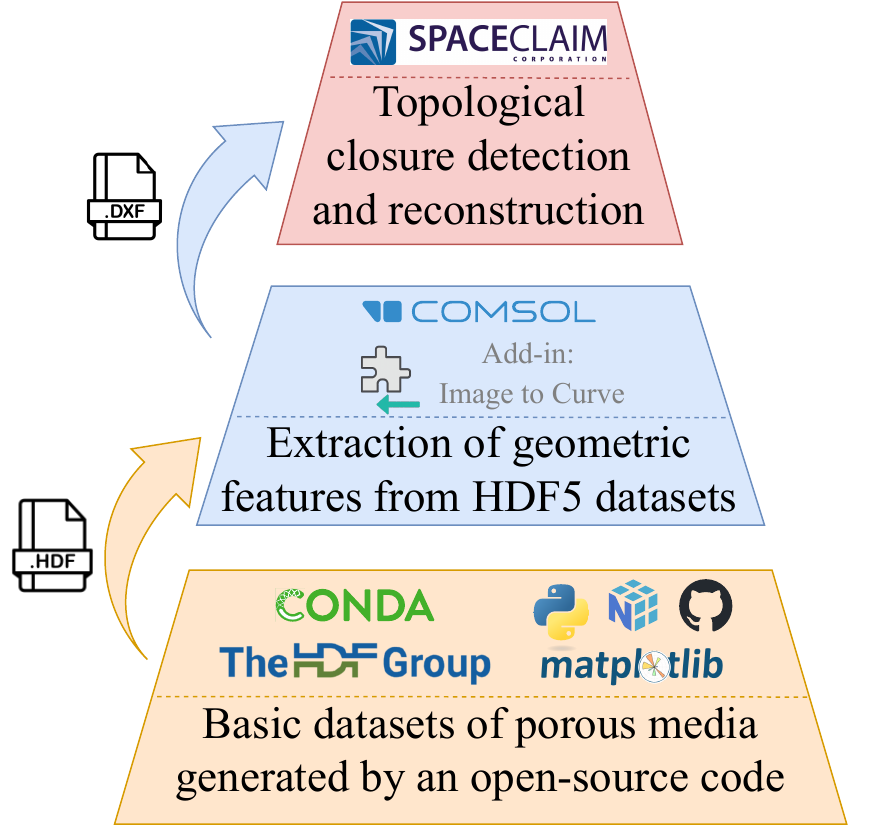}
    \caption{Workflow of the parameter-driven reconstruction of the in-wall CDPF porous media.}
    \label{fig:reconstruct workflow}
\end{figure}

Reconstructing porous media geometric models based on XRT or SEM images has several inherent limitations, such as the limited image resolution, potential artifacts introduced during image segmentation, and significant computational expense associated with 3-D reconstruction. Apart from these, the most critical drawback is that this approach is not suitable for parametric or systematic studies, which requires a series of porous structure with different geometric parameters. \cite{tahmasebi2012reconstruction,bodla20143d,mosser2017reconstruction,zhang20223d} Therefore, in this study, a parameter-driven approach is developed to reconstruct the in-wall CDPF porous structure. The workflow of the proposed method is illustrated in Fig.~\ref{fig:reconstruct workflow}.

The process can be divided into three main steps. 

\textbf{Step 1}: A basic dataset of porous media is generated by an open-source code provided by Dr.~Hannah Menke \cite{abdellatif2025benchmark}. General features of the code has been concluded in Algorithm~\ref{algorithm:porous media generator}. In addition to the three basic parameters (\texttt{rad}, \texttt{xdim}, \texttt{ydim}) which define the radius of a circle and the overall size of the porous media, the spatial anisotropy of the generated structure is achieved by controlling these parameters, namely \texttt{raddevmax}, \texttt{xdevmax}, \texttt{ydevmax}. In this study, we choose the following dimensionless parameter configurations: $\texttt{rad} = 8$, $\texttt{xdim} = 200$, $\texttt{ydim} = 800$, $\texttt{raddevmax} = 0.5$, $\texttt{xdevmax} = 0.4$, $\texttt{ydevmax} = 0.4$.~Using this set of parameters, the spatial heterogeneity of CDPF porous media \cite{lou2024novel} can be clearly represented.

\begin{center}
    \begin{minipage}{1.0\linewidth}
    \begin{algorithm}[H]
        \caption{General features of the porous media generator}
        \label{algorithm:porous media generator}
        \begin{algorithmic}
            \Require \texttt{rad}, \texttt{xdim}, \texttt{ydim}, \texttt{raddevmax}, \texttt{xdevmax}, \texttt{ydevmax}.
            \Ensure A \texttt{HDF5} file that consists of the coordinates, radii, binary image and metadata.
        \end{algorithmic}
        \begin{algorithmic}[1]
            \State Generate coordinates for an offset grid of circles by calling \texttt{generate\_offset\_grid()}.
            \State Render circles into a binary numpy array and compute porosity by calling \texttt{draw\_circles()}.
            \State Write coordinates and image data to an HDF5 file by calling \texttt{save\_to\_hdf5()}.
        \end{algorithmic}
    \end{algorithm}
    \end{minipage}
\end{center}

\textbf{Step 2}: The curve features are extracted from the previous HDF5 file using a COMSOL's add-in named Image-to-Curve \cite{luo2025microscopic}, with Gaussian filtering, contour thresholding, curve tolerance adjusting, and pixel refinement applied to enhance accuracy and eliminate noisy boundaries \cite{lou2024numerical}.

\textbf{Step 3}: SpaceClaim \cite{zhao2025research}, a pre-processing software, is used to check the topological closure of the curves from DXF file and to construct the corresponding solid geometry, which is surface in a 2-D model. 

\begin{figure}[ht]
    \centering
    \includegraphics[width=0.85\linewidth]{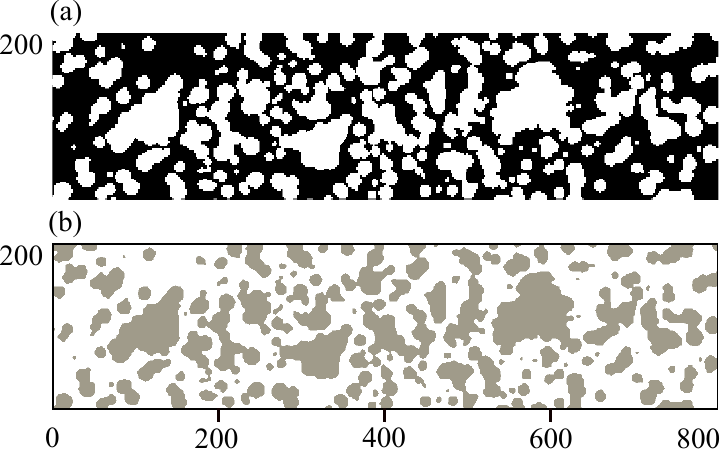}
    \caption{Results of the parameter-driven reconstruction of the in-wall CDPF porous media with a porosity of 0.56 and an effective grain size of~$2.07\times10^{-5}~\textrm{m}$: (a) Image of the porous structure extracted from the HDF5 file. (b) Final geometry reconstructed by SpaceClaim.}
    \label{fig:reconstruction porous structure}
\end{figure}

Fig.~\ref{fig:reconstruction porous structure} shows the result of the parameter-driven reconstruction of the in-wall CDPF porous media. The porosity of our model is 0.56 and the effective grain size is $2.07 \times 10^{-5}~\mathrm{m}$. The geometric heterogeneity of the CDPF porous structure is well represented.

\subsection{Development of the multiphysics mathematical model for the CDPF} 
\label{sec:development of the multiphysics mathematical model for the CDPF}

\subsubsection{Basic flow model}
\label{sec:basic flow model}
The conservation equations for an incompressible Newtonian fluid flow \cite{weller1998tensorial} can be described as Eqs.~(\ref{eq:continuity eqn}) and (\ref{eq:N-S eqn}),
\begin{equation}
    \nabla \cdot \boldsymbol{u} = 0, \label{eq:continuity eqn}
\end{equation}
\begin{equation}
    \frac{\partial \boldsymbol{u}}{\partial t} + \nabla \cdot (\boldsymbol{u} \boldsymbol{u}) = -\nabla \tilde{p} + \nabla \cdot (\nu_{\textrm{f}} \nabla \boldsymbol{u}) + \boldsymbol{g}, \label{eq:N-S eqn}
\end{equation}
where $\boldsymbol{u}$ represents the fluid velocity vector, $\tilde{p}$ denotes the kinematic pressure ($\tilde{p} = p / \rho_\textrm{f}$, where $\rho_\textrm{f}$ is the fluid density), $\nu_{\textrm{f}}$ is the kinematic viscosity ($\nu_\textrm{f} = \mu_\textrm{f} / \rho_\textrm{f}$), and $\boldsymbol{g}$ symbolizes the gravitational acceleration.

\subsubsection{Species transport model coupled with reaction kinetics}
\label{sec:species transport model coupled with reaction kinetics}
The gas mixture entering the CDPF consists of five species: $\textrm{O}_2$, $\textrm{NO}_2$, $\textrm{NO}$, $\textrm{CO}_2$ and $\textrm{N}_2$ \cite{kong2019numerical, duan2025effects}. The transport of each species can be modeled by Eq.~(\ref{eq:species transport eqn}) \cite{incropera1996fundamentals},
\begin{equation}
    \frac{\partial Y_i}{\partial t} + \nabla \cdot (\boldsymbol{u} Y_i) = \nabla \cdot (D_{\textrm{f},i} \nabla Y_i) + S_i, \label{eq:species transport eqn}
\end{equation}
where the subscript $i$ represents the $i$-th species, $Y$ is the mass fraction, $D_\textrm{f}$ is the mass diffusion coefficient, and $S$ is the source term.

\begin{figure}[ht]
    \centering
    \includegraphics[width=0.85\linewidth]{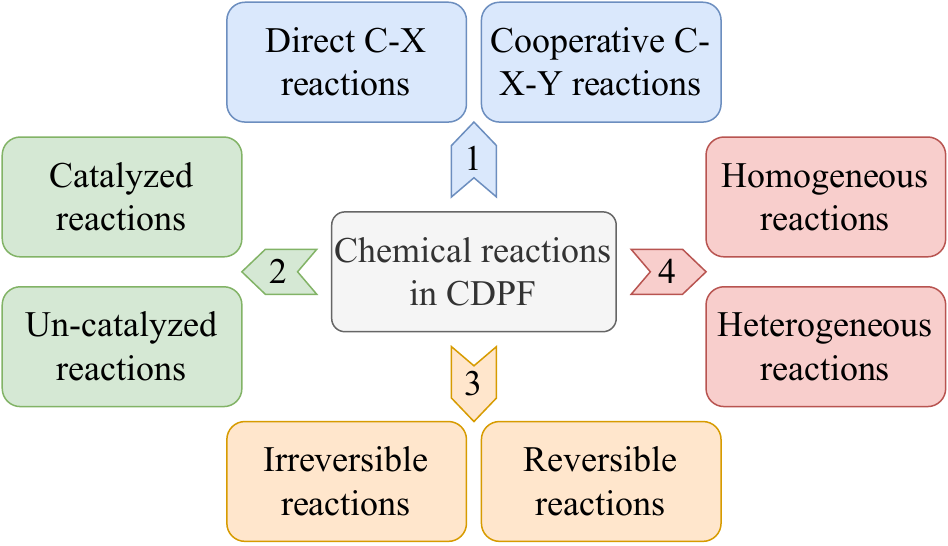}
    \caption{Classification of chemical reactions in CDPF.}
    \label{fig:classification of chemical reactions in CDPF}
\end{figure}

Eqs.~(\ref{eq:C-O2 pathway})-(\ref{eq:NO-O2-NO2 reaction}) summarize the main chemical reactions in CDPF \cite{tan2019modeling}. Based on the number and phase of the reactants, the presence or absence of a catalyst, and the reversibility of a reaction, we classify the above chemical reactions into four categories, as shown in Fig.~\ref{fig:classification of chemical reactions in CDPF}.

R1: Direct $\textrm{C}-\textrm{O}_2$ reaction path,
\begin{equation}
    \textrm{C(s)} + \textrm{O}_2(\textrm{g}) \rightarrow \textrm{CO}_2(\textrm{g}).
    \label{eq:C-O2 pathway}
\end{equation}

R2: Direct $\textrm{C}-\textrm{NO}_2$ reaction path,
\begin{equation}
    \textrm{C(s)} + 2\textrm{NO}_2(\textrm{g}) \rightarrow \textrm{CO}_2(\textrm{g})+2\textrm{NO}(\textrm{g}).
    \label{eq:C-NO2 pathway}
\end{equation}

R3: Cooperative $\textrm{C}-\textrm{O}_2-\textrm{NO}_2$ reaction path,
\begin{equation}
    \textrm{C(s)} + \textrm{NO}_2(\textrm{g}) + 0.5\textrm{O}_2(\textrm{g}) \rightarrow \textrm{CO}_2(\textrm{g})+\textrm{NO}(\textrm{g}).
    \label{eq:C-O2-NO2 pathway}
\end{equation}

R4: Reversible homogeneous reaction,
\begin{equation}
    \textrm{NO}(\textrm{g}) + 0.5\textrm{O}_2(\textrm{g}) \rightleftharpoons \textrm{NO}_2(\textrm{g}).
    \label{eq:NO-O2-NO2 reaction}
\end{equation}

In this study, we assume that all heterogeneous reactions (R1-R3) only occur at the CDPF substrate surface and there is no accumulation at the interface during reaction \cite{ou2022directional}. A Robin boundary condition \cite{mazumder2015numerical} Eq.~(\ref{eq:robin b.c. for species}) is then developed for describing the species balance at the interface,
\begin{equation}
    \rho_\textrm{f} D_{\textrm{f},i} \nabla Y_i \cdot \boldsymbol{n}+ \dot m_i = 0, 
    \label{eq:robin b.c. for species}
\end{equation}
where $\boldsymbol{n}$ is the surface normal vector, and $m_i$ is the mass production rate of species $i$ due to heterogeneous reactions, which can be modeled by Eq.~(\ref{eq:mass prod rate for heterogeneous}),
\begin{equation}
    \dot m_i = M_i \sum_{j=1}^{R} \nu_{i,j} v_{\textrm{r}, j},
    \label{eq:mass prod rate for heterogeneous}
\end{equation}
where the subscript $j$ represents the $j$-th reaction, $M$ is the molar mass, $\nu$ is the stoichiometric number, and $v_\textrm{r}$ is the reaction rate computed by the rate law, of which derivation can be found in Appendix.~\ref{app:rate laws derivation}.

For the homogeneous reaction R4, the mass production rate of species $i$ is incorporated into Eq.~(\ref{eq:species transport eqn}) as a source term, which is expressed as Eq.~(\ref{eq:source term for species trans eqn}),
\begin{equation}
    S_i = \dot m_i, \label{eq:source term for species trans eqn}
\end{equation}
and,
\begin{equation}
    \dot m_i = M_i \nu_i \left(v_{\textrm{r,forward}} - v_{\textrm{r, reverse}} \right),
    \label{eq:mass prod rate for homomgeneous}
\end{equation}
where $v_\textrm{r, forward}$ is the forward reaction rate, and $v_\textrm{r,reverse}$ is the reversed reaction rate. Details on Eq.~(\ref{eq:mass prod rate for homomgeneous}) can also be found in Appendix.~\ref{app:rate laws derivation}.

\subsubsection{Heat transfer model coupled with thermochemistry}
\label{sec:heat transfer model coupled with thermochemistry}
Soot oxidation is an exothermic process \cite{sharma2012experimental} and can be expressed using the energy equation \cite{incropera1996fundamentals} Eq.~(\ref{eq:energy eqn}) with applying the thermodynamic closure models in Sec.~\ref{sec:thermodynamic closures},
\begin{equation}
    \frac{\partial T}{\partial t} + \nabla \cdot (\boldsymbol{u} T) = \nabla \cdot (\alpha_\textrm{f} \nabla T) + Q_{T},
    \label{eq:energy eqn}
\end{equation}
where $T$ is the fluid temperature, $\alpha_\textrm{f}$ represents the fluid thermal diffusivity ($\alpha_\textrm{f} = k_{\textrm{f}} \rho_{\textrm{f}}^{-1}C_{p,\textrm{f}}^{-1}$, where $k_{\textrm{f}}$ is the thermal conductivity and $C_{p,\textrm{f}}$ is the specific heat capacity of fluid), and $Q_T$ is expressed as Eq.~(\ref{eq:Q_T in energy eqn}),
\begin{equation}
    Q_T = \frac{1}{\rho_\textrm{f} C_{p,\textrm{f}}} \left( \dot{q}_{V} + \rho_\textrm{f} T \frac{D C_{p,\textrm{f}}}{D t} \right),
    \label{eq:Q_T in energy eqn}
\end{equation}
where $\dot q_V$ is the rate of heat production.

For heterogeneous reactions (R1-R3), a heat flux boundary condition \cite{patankar2018numerical} Eq.~(\ref{eq:heat flux b.c}) is set at the interface,
\begin{equation}
    q^{\prime \prime} = \sum_{j=1}^{R} \left(-v_{\textrm{r},j} \Delta_{\textrm{r}}H^{\plimsoll}_j \right) - k_\textrm{f} \nabla T \cdot \boldsymbol{n},
    \label{eq:heat flux b.c}
\end{equation}
where the subscript $j$ denotes the $j$-th heterogeneous chemical reaction, $\Delta_{\textrm{r}}H^{\plimsoll}$ is the temperature-dependent standard reaction enthalpy change, which can be calculated by the Kirchhoff's law \cite{atkins2023atkins} Eq.~(\ref{eq:Kirchhoff's law}),
\begin{equation}
    \Delta_{\textrm{r}}H^{\plimsoll}(T) = \Delta_{\textrm{r}}H^{\plimsoll}(T_\textrm{ref}) + \int_{T_{\textrm{ref}}}^{T} \Delta_{\textrm{r}} C_p^{\plimsoll} dT,
    \label{eq:Kirchhoff's law}
\end{equation}
where $T_\textrm{ref}$ is the reference temperature ($293.15~\mathrm{K}$), and $\Delta_{\textrm{r}} C_p^{\plimsoll}$ is calculated by Eq.~(\ref{eq:sensible enthalpy}),
\begin{equation}
    \Delta_{\textrm{r}} C_p^{\plimsoll} = \sum_{i=1}^{N} \nu_i C_{p, \textrm{mol},i}^{\plimsoll},
    \label{eq:sensible enthalpy}
\end{equation}
where $C_{p, \textrm{mol}}$ represents the molar heat capacity at constant pressure, which can be approximated by the Shomate Equation \cite{linstrom2001nist}.

For the heat release of R4, the same approach as discussed in Sec.~\ref{sec:species transport model coupled with reaction kinetics} is adopted, where the released heat is directly incorporated as a source term in Eq.~(\ref{eq:energy eqn}),
\begin{equation}
    \dot q_V = -\left(v_{\textrm{r,forward}} - v_{\textrm{r, reverse}} \right) \Delta_\textrm{r} H^\plimsoll.
    \label{eq:source term for energy eqn}
\end{equation}

\subsubsection{Thermodynamic closures}
\label{sec:thermodynamic closures}
Closure of the governing equations discussed above is achieved by specifying the thermodynamic properties of the gas mixture. Table.~\ref{tab:thermo closures} concludes the thermodynamic closure models adopted in this study. Details on these models can be found in \cite{welahettige2016comparison}.
\begin{table}[ht]
    \caption{Closure models for physical properties of the gas mixture.}
    \label{tab:thermo closures}
    \begin{ruledtabular}
    \begin{tabular}{>{\raggedright\arraybackslash}p{0.36\linewidth} >{\raggedright\arraybackslash}p{0.58\linewidth}}
        Physical property & Thermodynamic closure model \\
        \hline
        Density,~$\rho_\textrm{f}$ & Incompressible ideal gas model \\
        Specific heat,~$c_{p,\textrm{f}}$ & Mass fraction averaged mixing model \\
        Thermal conductivity,~$k_\textrm{f}$ & Kinetic theory based mixing model \\
        Dynamic viscosity,~$\mu_\textrm{f}$ & Kinetic theory based mixing model \\
        Mass diffusivity,~$D_{\textrm{f},i}$ & Constant dilute approximation, $7.3\times10^{-5} ~\mathrm{m^2.s^{-1}}$
    \end{tabular}
    \end{ruledtabular}
\end{table}

\subsubsection{Particle transport and deposition model}
\label{sec:particle transport and deposition model}
In this study, the dilute flow assumption is adopted and the coupling between the fluid and dispersed phase is assumed to be one-way \cite{parker2022statistically}. Also, the rotational motion is not considered since the momentum inertia for the ultra-fine soot particulate is negligible. The linear momentum equation Eq.~(\ref{eq:particle motion eqn}) for each particle with a mass $m_\textrm{p}$ can be derived through the basic Newton's second law,
\begin{equation}
    m_{\textrm{p}} \frac{d \boldsymbol{u}_{\textrm{p}}}{d t} = \boldsymbol{F}_{\textrm{d}} + \boldsymbol{F}_{\textrm{b-w}}  + \boldsymbol{F}_{\textrm{b}} + \boldsymbol{F}_{\textrm{t}} + \boldsymbol{F}_{\textrm{v}} + \boldsymbol{F}_{\textrm{p}} + \boldsymbol{F}_{\textrm{l}},
    \label{eq:particle motion eqn}
\end{equation}
where $\boldsymbol{u}_{\textrm{p}}$ is the particle velocity. A total of seven forces are considered, namely the drag force $\boldsymbol{F}_\textrm{d}$, the buoyancy and weight $\boldsymbol{F}_\textrm{b-w}$, the Brownian force $\boldsymbol{F}_\textrm{b}$, the thermophoretic force $\boldsymbol{F}_\textrm{t}$, the virtual mass force $\boldsymbol{F}_\textrm{v}$, the pressure gradient force $\boldsymbol{F}_\textrm{p}$, and the Saffman's lift $\boldsymbol{F}_\textrm{l}$. The definitions of them can be found in Appendix.~\ref{app:forces acting on a soot particle}.

Two conditions are required for particle deposition: physical contact with the surface and adhesion to it \cite{haugen2010particle}. Based on these, the deposition model for ultra-fine soot particles is then developed. In brief, deposition is determined by comparing the particle's critical sticking velocity with its actual normal impact velocity: a particle deposits if the normal impact velocity is lower than the critical value; otherwise, it rebounds.

Previous studies \cite{wang1991filtration,fang2022new,wang2024numerical} have shown that at micrometer scale, adhesion due to van der Waals force becomes non-negligible and acts to hold the particle to the surface. For nano-scale particles, an analytical solution, pioneered by Wang and Kasper \cite{wang1991filtration}, for the critical sticking velocity $u_\textrm{p,cr}$ based on the Bradley and Hamaker (B–H) theory is adopted, which is expressed as Eq.~(\ref{eq:critical sticking velocity}),
\begin{equation}
    u_\textrm{p,cr} = \sqrt{\frac{A_\textrm{h}}{\pi \rho_\textrm{p} Z_\textrm{sep} d_\textrm{p}^2}},
    \label{eq:critical sticking velocity}
\end{equation}
where $A_\textrm{h}$ is the Hamaker constant, and $Z_\textrm{sep}$ is the minimum separation distance ($0.4~\mathrm{nm}$). The Hamaker constant can be derived by the effective surface energy $\gamma_\textrm{eff}$, as shown in Eq.~(\ref{eq:Hamaker constant}),
\begin{equation}
    A_\textrm{h} = 24 \pi \gamma_\textrm{eff} Z_{\textrm{sep},0}^2,
    \label{eq:Hamaker constant}
\end{equation}
where $Z_{\textrm{sep},0}$~is the cut-off separation distance (0.165~$\mathrm{nm}$), and the effective surface energy is the geometric mean of the particle ($\gamma_1=0.095~\textrm{J.m}^{-2}$) and impact-wall ($\gamma_2=2.561~\textrm{J.m}^{-2}$) surface energy \cite{wei2018graphite, gu2018structure}.

Considering the energy loss during the impaction, a restitution coefficient \cite{ahmad2016impact} $e$ is then applied to accurately model the particle rebound process. Its definition is given in Eq.~(\ref{eq:restitution coeff}), 
\begin{equation}
    e = \sqrt{1-\frac{u_\textrm{p,cr}}{u_\textrm{p,i}}},
    \label{eq:restitution coeff}
\end{equation}
where $u_\textrm{p,i}$ is the particle incident velocity.

\subsection{Numerical implementation}
\label{sec:numerical implementation}

\subsubsection{Computational domain}
\label{sec:computational domain}
\begin{figure}[ht]
    \centering
    \includegraphics[width=1.0\linewidth]{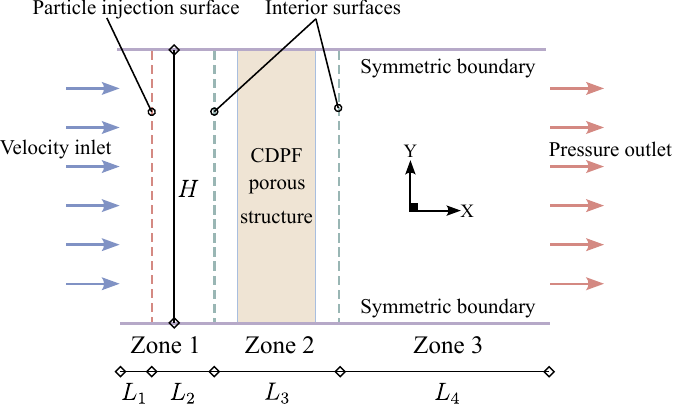}
    \caption{Schematic diagram of the computational domain for the CDPF porous structure.}
    \label{fig:computational domain cdpf}
\end{figure}

Fig.~\ref{fig:computational domain cdpf} shows the computational domain used for the numerical simulation. Upstream and downstream regions are extended adjacent to the porous geometry to ensure fully developed flow and reflect the actual CDPF channel structure. Details on the geometric parameters can be found in Table.~\ref{tab:geometric param of the cdpf computational domain}. A two-dimensional unstructured mesh with a total of \num{659073} cells is built using ICEM. Details on the mesh setup and independence check will be discussed in Sec.~\ref{sec:discretization error analysis and model validation}.

\begin{table}[ht]
    \caption{Geometric parameters of the computational domain.}
    \label{tab:geometric param of the cdpf computational domain}
    \begin{ruledtabular}
    \begin{tabular}{lll}
        Parameter & Value & Unit \\
        \hline
        $L_1$ & $4\times10^{-5}$ & $\mathrm{m}$ \\
        $L_2$ & $1.6\times10^{-4}$ & $\mathrm{m}$ \\
        $L_3$ & $4\times10^{-4}$ & $\mathrm{m}$ \\
        $L_4$ & $4\times10^{-4}$ & $\mathrm{m}$ \\
        $H$ & $8\times10^{-4}$ & $\mathrm{m}$ \\
    \end{tabular}
    \end{ruledtabular}
\end{table}

\subsubsection{Solution algorithm}
\label{sec:solution algorithm}
In this study, ANSYS Fluent 2402, a commercial CFD code based on FVM, is selected as the numerical solver \cite{welahettige2016comparison}. To incorporate the mathematical model derived in Sec.~\ref{sec:development of the multiphysics mathematical model for the CDPF}, several user-defined functions (UDFs) and user-defined memories (UDMs) were programmed using \texttt{C}. The definitions and usages of them are summarized in Table.~\ref{tab:udfs and udms} in Appendix.~\ref{app:udfs and udms}. 

Fig.~\ref{fig:solution algorithm for cdpf} demonstrates the solution algorithm for our model. An Eulerian-Lagrangian framework is adopted. The Eulerian approach is employed to solve the fluid flow (left side), while the Lagrangian method is applied for the particle tracking (right side). Note that all UDFs and UDMs must be loaded in advance before the computation begins. To accelerate the transient computing, we first performed a steady-state simulation and used its results as the initial field for the transient simulation through interpolation. As mentioned above, the coupling between the fluid and particles is one-way. Back into the algorithm, once the solution for the fluid flow is converged in the current time-step, the data of the flow field will be passed to the particle solver, which is the discrete phase model (DPM) solver in Fluent \cite{parker2022statistically}. Considering the unsteadiness in the continuous phase, a transient particle tracking method is adopted. The maximum number of tracking steps is set to be $5\times10^4$ in case a particle gets stuck in a recirculation zone. To increase the accuracy of tracking, the step length factor is set to be 5, and a high-resolution tracking is enabled. The tracking scheme is automated, that is, a low-order implicit scheme is adopted when the convergence is poor; otherwise, a high-order trapezoidal scheme is chosen. Once the particle tracking procedure is completed in a fluid-flow time-step, the solution will be saved and the solver for the continuous phase will proceed to the next time step, and this cycle continues until the specified end time.

When solving the governing equations for the continuous phase, a central-difference scheme is applied to the diffusion terms, while all convection terms are discretized through a second-order upwind scheme. To address the "checkerboard" pressure oscillation problem, a momentum-based Rhie-Chow interpolation \cite{bartholomew2018unified} is utilized to evaluate the pressure gradient at cell faces. For the unsteady term in the transient simulation, a second order implicit numerical scheme is adopted, which alleviates the Courant–Friedrichs–Lewy (CFL) restriction \cite{de2013courant}. For the pressure–velocity coupling, a segregated solver is employed. The Semi-Implicit Method for Pressure-Linked Equations (SIMPLE) algorithm \cite{patankar2018numerical} is used for the steady-state simulation to obtain the initial flow field, whereas the Pressure-Implicit with Splitting of Operators (PISO) algorithm \cite{issa1986solution} is adopted for the transient computations. The iteration is terminated once the residuals drop below $10^{-6}$, which is taken as the convergence criterion. The entire transient simulation is carried out for a duration of $2~\mathrm{s}$ with a time step of $0.001~\mathrm{s}$, and particle injection begins at $0.6~\mathrm{s}$ and lasts for $0.2~\mathrm{s}$ within the computational domain.
\begin{figure*}
    \centering
    \includegraphics[width=0.8\linewidth]{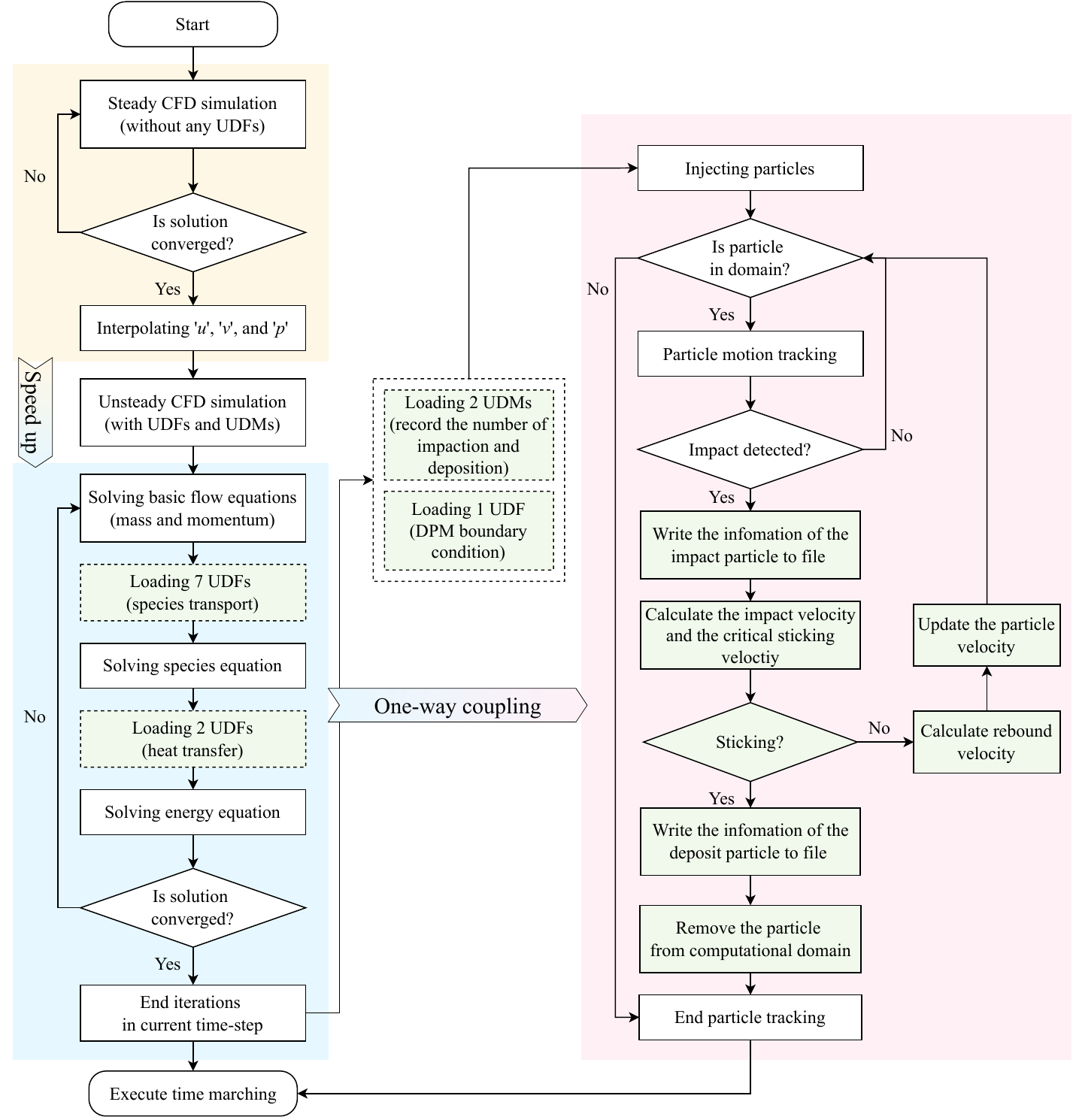}
    \caption{Flowchart of the numerical solution algorithm for the CDPF mathematical model.}
    \label{fig:solution algorithm for cdpf}
\end{figure*}

\subsubsection{Simulation details}
\label{sec:simulation details}
To evaluate the proposed model's capability in capturing species and heat transport during CDPF regeneration, an operating condition is selected with an inlet temperature of $600~\mathrm{K}$ and the mass fractions of 0.1, 0.0003, 0.0003, 0.1, and 0.7994 for $\textrm{O}_2$, $\textrm{NO}_2$, $\textrm{NO}$, $\textrm{CO}_2$ and $\textrm{N}_2$, respectively \cite{huang2025simulation}. In CDPF porous media, the flow typically exhibits a Reynolds number less than 1 \cite{huang2024study}. Based on this characteristic, an inlet Reynolds number $Re_H = u_0 H / \nu_\textrm{f}$ of 0.5 is chosen. Combining the density and viscosity results computed from the closure model for the fluid properties, the inlet velocity is then determined. 

Although the DPM model adopts a particle-in-cell (parcel) tracking strategy \cite{song2018sub} to improve computational efficiency, the total number of particles in our simulation is relatively small. Consequently, each parcel is defined to contain only a single particle in the present model. At each time step, approximately $100$ inert particles with a diameter of $50~\mathrm{nm}$ are injected. Their initial velocity and temperature are set to be identical to those of the inlet mixture. The density, specific heat capacity, and thermal conductivity for a soot particle are $1.77\times10^{3}~\mathrm{kg.m^{-3}}$, $1.58\times10^{3}~\mathrm{J.kg^{-1}.K^{-1}}$, and $0.6~\mathrm{W.m^{-1}.K^{-1}}$, respectively.

It is also worth noting that, strictly speaking, Eq.~(\ref{eq:robin b.c. for species}) corresponds to a Robin boundary condition, which is not directly supported in Fluent. Therefore, a fixed-gradient boundary condition is adopted, with the interfacial gradient specified according to the mass production rate of the species in this study.

\section{Benchmark-based code verification}
\label{sec:benchmark based code verification}
In this section, a series of tests on the UDF codes developed based on the mathematical model discussed above has been conducted. All verifications are conducted based on a classical two-dimensional circular cylinder flow problem. In our first test case, we examine the mass transport on the reacting surface of the circular cylinder within an unsteady pure-diffusion framework. Then, we shift our focus to the solution of this problem under a convection-diffusion scenario. Finally, by varying the Stokes numbers $St$ and the Reynolds numbers $Re$, of which the definitions will be discussed later, we complete the verification concerning the particle-wall impaction behaviors.

\subsection{Surface reaction kinetics}
\label{sec:surface reaction kinetics}

\subsubsection{Diffusion problem}
\label{sec:diffusion problem}
The diffusion phenomena of a reactive species towards the 2-D cylinder surface is examined first to assess the capability of the proposed model in describing problems involving heterogeneous reactions. A 3-D version of this benchmark has been successfully solved using both analytical and numerical approaches in previous studies \cite{lu2018direct,ou2022directional}. The geometric parameters of the computational domain remain the same as those in \cite{lu2018direct}, except for those in the $z$-direction, which does not exist in the current 2-D space. Specifically, a circular cylinder with a diameter $d_{\textrm{cyl}}$ is placed at the centroid of a square domain with a length $L$. The cylinder is then surrounded by a quiescent fluid with an initial concentration $c_{\textrm{f},0}$, which will be gradually consumed by a first-order irreversible chemical reaction at the cylinder surface, in exact accordance with Eq.(\ref{eq:robin b.c. for species}). For simplicity, only one species is considered in the fluid phase. Hence, a user-defined scalar (UDS) transport equation including the unsteady and diffusion terms is constructed to avoid the requirement of having at least two species in order to activate the built-in species transport equation in Fluent. The parameters used in this simulation have been listed in Table.~\ref{tab:parameters of the diffusion problem}.
\begin{table}[ht]
    \caption{Parameters used for the simulation of the surface reaction within an unsteady pure-diffusion framework.} 
    \label{tab:parameters of the diffusion problem}
    \begin{ruledtabular}
    \begin{tabular}{ll >{\raggedright\arraybackslash}p{0.25\linewidth} l}
        Parameter & Symbol & Value & Unit \\
        \hline
        Domain size & $L$ & $0.04$ & $\mathrm{m}$ \\
        Cylinder diameter & $d_{\textrm{cyl}}$ & $0.005$ & $\mathrm{m}$ \\
        Mesh size & $\Delta h$ & $2.0\times10^{-5}~\text{-}~\newline 2.5\times10^{-4}$ & $\mathrm{m}$ \\
        Time step & $\Delta t$ & $1\times10^{-5}$ & $\mathrm{s}$ \\
        Density & $\rho_\textrm{f}$ & $1$ & $\mathrm{kg.m^{-3}}$ \\
        Mass diffusivity & $D_\textrm{f}$ & $2\times10^{-5}$ & $\mathrm{m^2.s^{-1}}$ \\
        Initial concentration & $c_{\textrm{f},0}$ & $10$ & $\mathrm{mol.m^{-3}}$ \\ 
    \end{tabular}
    \end{ruledtabular}
\end{table}

Three dimensionless groups are selected for further analysis, namely the diffusion Damköhler number ($Da_{\textrm{diff}}$), Fourier number ($Fo$), and Sherwood number ($Sh$). The definitions of them are listed through Eqs.~(\ref{eq:diffusion Damkohler (verfi.case.01)})-(\ref{eq:Sherwood (verfi.case.01)}),
\begin{equation}
    Da = \frac{k_{\textrm{r}} d_{\textrm{cyl}}}{2 D_\textrm{f}}, \label{eq:diffusion Damkohler (verfi.case.01)}
\end{equation}
\begin{equation}
    Fo = \frac{4 D_\textrm{f} t}{d_{\textrm{cyl}}^2}, \label{eq:Fourier (verfi.case.01)}
\end{equation}
\begin{equation}
    Sh = \frac{\iint_{A_\textrm{cyl}} \left( \nabla c_\textrm{f} \cdot \boldsymbol{n} \right) \,dA}{A_\textrm{cyl} \left( c_{\textrm{f}, 0} - \langle c_{\textrm{f}} \rangle_A \right)} \frac{d_\textrm{cyl}}{D_\textrm{f}}, \label{eq:Sherwood (verfi.case.01)}
\end{equation}
where $A_\textrm{cyl}$ is the surface area of the cylinder, and $\langle c_\textrm{f} \rangle_A$ is the area-weighted averaged of species concentration at cylinder surface. The calculation of the Sherwood number is automatically done by a post-processing UDF macro named \texttt{DEFINE\_REPORT\_DEFINITION\_FN}. Five reaction rates, corresponding to $Da=0.01$, $0.1$, $1$, $10$, and $100$, are applied in this verification. The results are then compared with an analytical solution derived by \cite{lu2018direct}.

\begin{figure}[ht]
    \centering
    \includegraphics[width=0.9\linewidth]{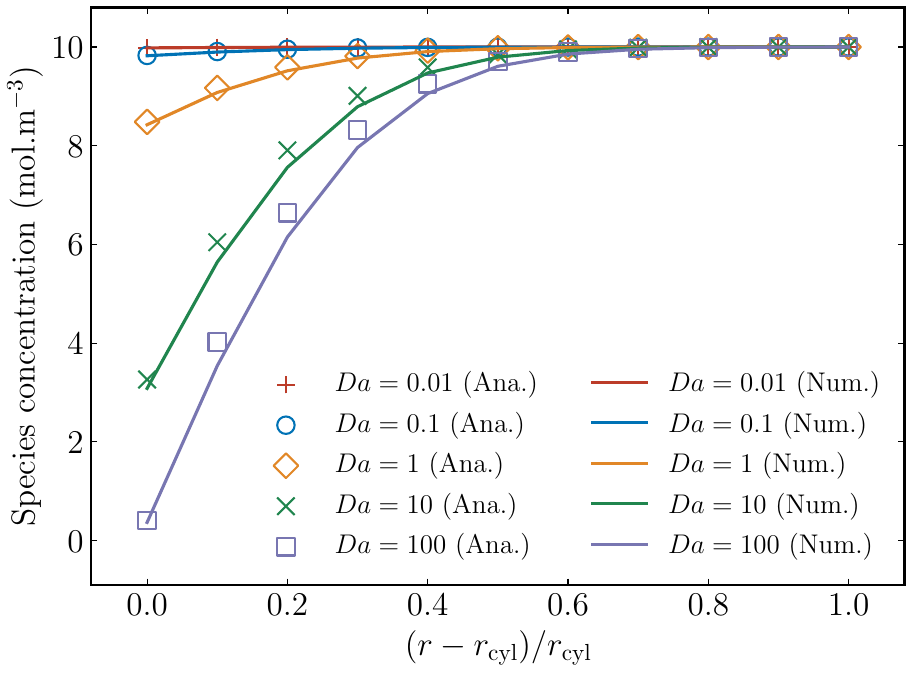}
    \caption{Spatial distributions of species concentration from analytical (Ana.) and numerical (Num.) results at different Damköhler numbers for a Fourier number of 0.032.}
    \label{fig:cf profile diffusion problem}
\end{figure}

\begin{figure}[ht]
    \centering
    \includegraphics[width=0.9\linewidth]{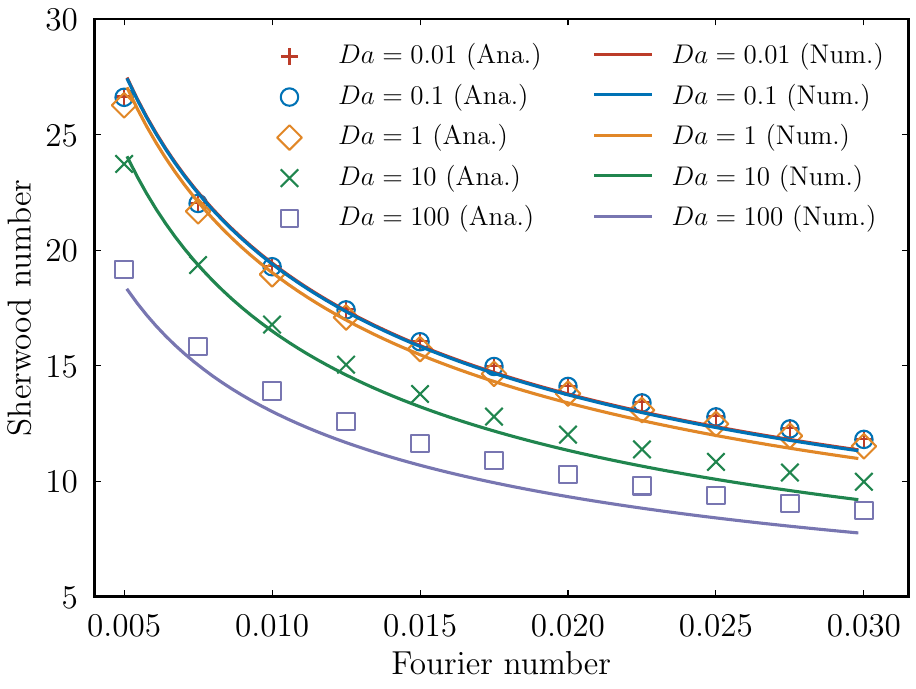}
    \caption{Temporal distribution of the Sherwood number from analytical (Ana.) and numerical (Num.) results at different Damköhler numbers.}
    \label{fig:sherwood diffusion problem}
\end{figure}

Fig.~\ref{fig:cf profile diffusion problem} illustrates the variations in species concentration near the cylinder surface at different Damköhler numbers when $Fo=0.032$. The simulation result exhibits good agreement with the analytical solution. As expected, at small Damköhler numbers the surface reaction is relatively weak, resulting in a more gradual concentration profile. With increasing Damköhler number, the reaction rate becomes faster, leading to a noticeably steeper concentration gradient. The temporal evolution of the Sherwood number under different Damköhler numbers is presented in Fig.~\ref{fig:sherwood diffusion problem}. Although some deviation appears at large Damköhler numbers as time progresses, good agreement is observed between the numerical result and the analytical solution. This discrepancy may stem from the fact that the analytical solution is derived from a three-dimensional model in which the reactive surface is assumed to be spherical.

\subsubsection{Convection-diffusion problem}
\label{sec:convection-diffusion problem}
All transport phenomena in fluid flow are governed by the combined effects of diffusion and convection \cite{versteeg2007introduction}. In the following part, an additional convection term is added to the previous UDS equation to test the performance of our UDF code in an actual flow problem. The geometry used in this part is the same as the previous one except that the cylinder is positioned $0.01~\mathrm{m}$ downstream from the inlet considering the potential flow structure changes behind the cylinder caused by various flow regimes. Table~\ref{tab:parameters of the conv-diff problem} summarizes the parameters used in this case. 
\begin{table}[ht]
    \caption{Parameters used for the simulation of the surface reaction within a convection-diffusion framework.}
    \label{tab:parameters of the conv-diff problem}
    \begin{ruledtabular}
        \begin{tabular}{ll >{\raggedright\arraybackslash}p{0.2\linewidth} l}
            Parameter & Symbol & Value & Unit \\
            \hline
            Domain size & $L$ & $0.04$ & $\mathrm{m}$ \\
            Cylinder diameter & $d_{\textrm{cyl}}$ & $0.005$ & $\mathrm{m}$ \\
            Mesh size & $\Delta h$ & $1.5\times10^{-5}~\text{-}~\newline 2.0\times10^{-4}$ & $\mathrm{m}$ \\
            Time step & $\Delta t$ & $1\times10^{-5}~\text{-}~\newline 1\times10^{-4}$ & $\mathrm{s}$ \\
            Species density & $\rho_\textrm{f}$ & $1$ & $\mathrm{kg.m^{-3}}$ \\
            Species viscosity & $\mu_\textrm{f}$ & $2\times10^{-5}$ & $\mathrm{Pa.s}$ \\
            Species diffusivity & $D_\textrm{f}$ & $2\times10^{-5}$ & $\mathrm{m^2.s^{-1}}$ \\
            Inlet concentration & $c_{\textrm{f},0}$ & $10$ & $\mathrm{mol.m^{-3}}$
        \end{tabular}
    \end{ruledtabular}
\end{table}

Two additional dimensionless groups are introduced, namely the cylinder Reynolds number $Re_{\textrm{cyl}}$ and the Schmidt number $Sc$, the definitions of them are shown as follows,
\begin{equation}
    Re_{\textrm{cyl}} = \frac{\rho_\textrm{f} u_0 d_{\textrm{cyl}}}{\mu_{\textrm{f}}}, \label{eq:Reynolds number used in conv-diff benchmark}
\end{equation}
\begin{equation}
    Sc = \frac{\mu_\textrm{f}}{\rho_\textrm{f} D_\textrm{f}}, \label{eq:Schmidt number used in conv-diff benchmark}
\end{equation}
where $u_0$ indicates the inlet velocity. An empirical relationship, expressed as Eq.~(\ref{eq:Hilpert correlation}), between $Sh$, $Re_\textrm{cyl}$, and $Sc$, discovered by \cite{hilpert1933warmeabgabe}, is then utilized to validate the results from the current simulation,
\begin{equation}
    Sh = \frac{k_\textrm{c} d_\textrm{cyl}}{D_\textrm{f}} = C Re_s^m Sc^{1/3}, \label{eq:Hilpert correlation}
\end{equation}
where $C$ and $m$ are the constants that can be found in \cite{hilpert1933warmeabgabe}, and $k_\textrm{c}$ is the external mass transfer coefficient, which can be further used to compute the overall effective mass transfer coefficient $k_\textrm{eff}$ through Eq.~(\ref{eq:effective mass transfer coefficient}),
\begin{equation}
    k_\textrm{eff} = \frac{k_\textrm{r} k_\textrm{c}}{k_\textrm{r} + k_\textrm{c}}. \label{eq:effective mass transfer coefficient}
\end{equation}

\begin{figure*}
    \centering
    \includegraphics[width=0.75\linewidth]{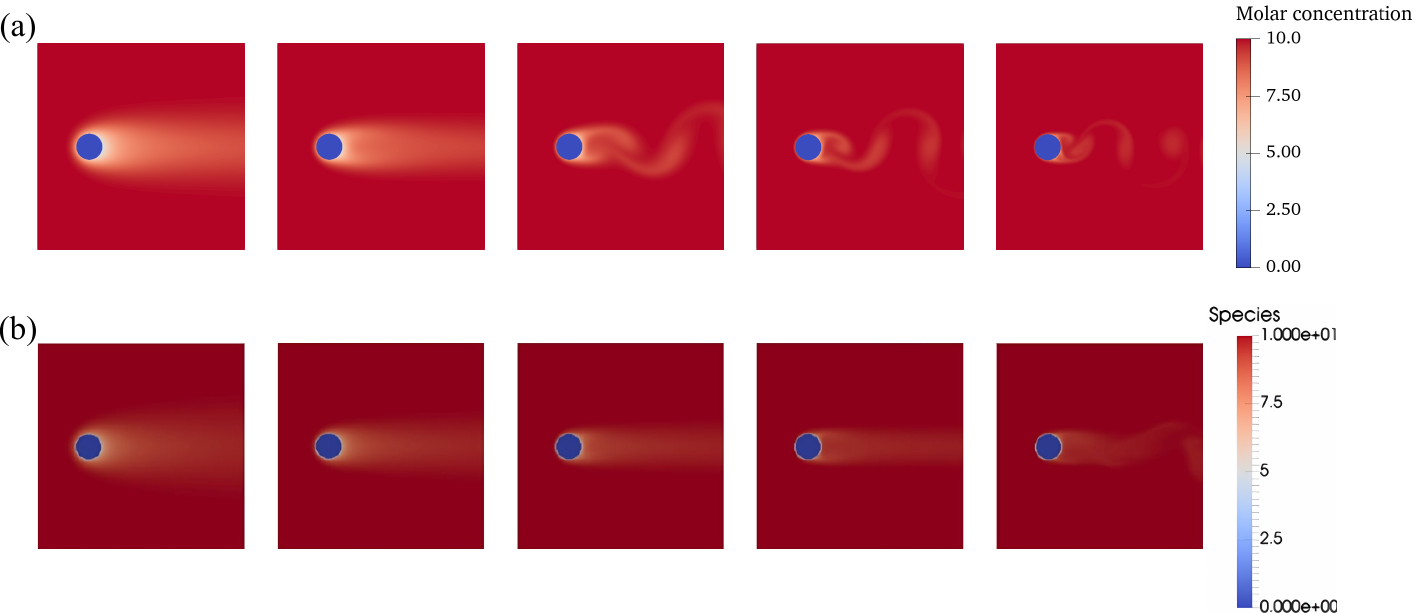}
    \caption{Snapshots of species concentration distribution under different cylinder Reynolds numbers (left to right: 20, 50, 100, 200 and 400) at a fixed Damköhler number (1): (a) Present study; (b) Ghost-cell based IBM \cite{lu2018direct}. J. Lu, S. Das, E. Peters, and J. Kuipers, Chemical Engineering Science 176, 1–18 (2018); licensed under a Creative Commons Attribution (CC BY) license.}
    \label{fig:snapshot molar concentration}
\end{figure*}

Five cylinder Reynolds numbers (20, 50, 100, 200, and 400) and one Damköhler number (1) are selected as the non-dimensional operating parameters for this simulation. The results of the species concentration distribution computed using our UDF code are compared with a previous study using the ghost-cell based immersed boundary method (IBM) \cite{lu2018direct}, as demonstrated in Fig.~\ref{fig:snapshot molar concentration}. Meanwhile, the overall effective mass transfer coefficients obtained from the present simulation are compared with those computed through Eq.~(\ref{eq:Hilpert correlation}) and (\ref{eq:effective mass transfer coefficient}). As illustrated in Table~\ref{tab:overall effective mass transfer coefficients},  a good level of consistency is achieved between them.

\begin{table}[ht]
    \caption{The overall effective mass transfer coefficients under five cylinder Reynolds numbers at $Da=1$.}
    \label{tab:overall effective mass transfer coefficients}
    \begin{ruledtabular}
    \begin{tabular}{>{\raggedright\arraybackslash}p{0.15\linewidth} >{\raggedright\arraybackslash}p{0.25\linewidth} >{\raggedright\arraybackslash}p{0.28\linewidth} >{\raggedright\arraybackslash}p{0.2\linewidth} }
        $Re_\textrm{cyl}$ & \multicolumn{2}{l}{Effective mass transport coefficient $k_\textrm{eff}$} & Error \\
        \cline{2-3}
          & Present study & Empirical formula &  \\ 
        \hline
        $20$ & $4.89 \times 10^{-3}$ & $4.74 \times 10^{-3}$ & $3.25\%$ \\
        $50$ & $5.49 \times 10^{-3}$ & $5.46 \times 10^{-3}$ & $0.59\%$ \\
        $100$ & $6.05 \times 10^{-3}$ & $5.96 \times 10^{-3}$ & $1.43\%$ \\
        $200$ & $6.52 \times 10^{-3}$ & $6.41 \times 10^{-3}$ & $1.61\%$ \\
        $400$ & $6.89 \times 10^{-3}$ & $6.78 \times 10^{-3}$ & $1.50\%$ \\
    \end{tabular}
    \end{ruledtabular}
\end{table}

\subsection{Particle-wall interaction}
\label{sec:particle-wall interaction}
Due to the uniqueness and complexity of the particle-wall interaction, in this part, we restrict ourselves to the inertial impaction only and the following sticking process is not considered. Therefore, only the particle drag force will be activated in the Fluent DPM solver and the corresponding UDF code mainly focuses on the message-passing and data output under parallelization.

\begin{figure}[ht]
    \centering
    \includegraphics[width=1.0\linewidth]{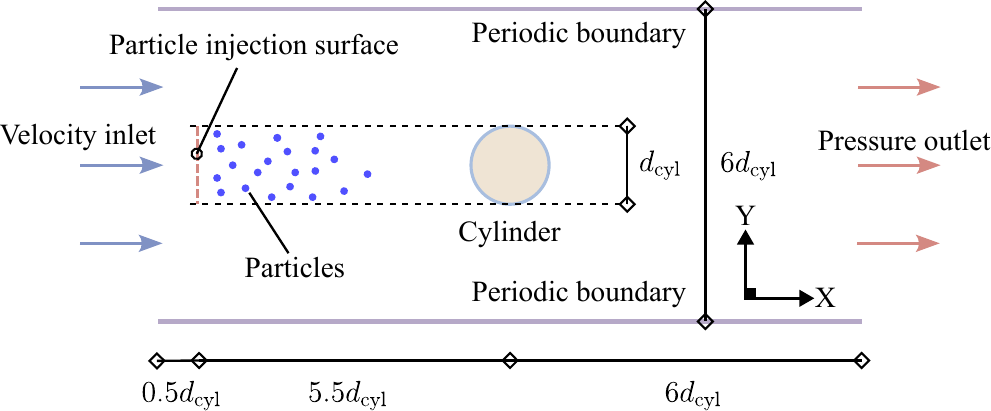}
    \caption{Schematic diagram of the computational domain for the code verification on particle-wall interaction.}
    \label{fig:computational domain benchmark}
\end{figure}

Still, a 2-D laminar flow past a circular cylinder is utilized as a building block in this verification. The computational domain and several boundary conditions are described in Fig.~\ref{fig:computational domain benchmark}. A series of cases under different operating conditions have been designed to verify our discrete phase model and the UDF code. For better explanation, apart from the cylinder Reynolds number ($Re_\textrm{cyl}$), a new dimensionless group called the Stokes number ($St$) is introduced. The definition of $St$ can be found in \cite{haugen2010particle}. Details on mesh setup can be referred to \cite{bouhairie2007two}. The parameters used in this simulation are concluded in Table~\ref{tab:parameter for particle benchmark}.

\begin{table}[ht]
    \caption{Parameters used for simulations of particle-wall interaction problem. \label{tab:parameter for particle benchmark}}
    \begin{ruledtabular}
    \begin{tabular}{ll >{\raggedright\arraybackslash}p{0.25\linewidth} l}
        Parameter & Symbol & Value & Unit \\
        \hline
        Cylinder diameter & $d_\textrm{cyl}$ & 0.005 & $\mathrm{m}$ \\
        Density ratio & $\rho_\textrm{p} / \rho_\textrm{f}$ & $1\times10^3$ & - \\
        Fluid viscosity & $\mu_\textrm{f}$ & $2\times10^{-5}$ & $\mathrm{Pa.s}$ \\
        Cylinder Reynolds number & $Re_\textrm{cyl}$ & 20, 100 & - \\
        Stokes number & $St$ & 0.01, 0.05, 0.1, 0.5, 1, 5, 10 & - \\
        Total injection time & $t_{\textrm{tot,inject}}$ & $0.5$ & $\mathrm{s}$ \\
        Total number of particle & $N_\textrm{tot}$ & $2.5\times10^{4}$ & - \\
        Time step & $\Delta t$ & $2\times10^{-4}$ & $\mathrm{s}$
    \end{tabular}
    \end{ruledtabular}
\end{table}

\begin{figure*}
    \centering
    \includegraphics[width=0.65\linewidth]{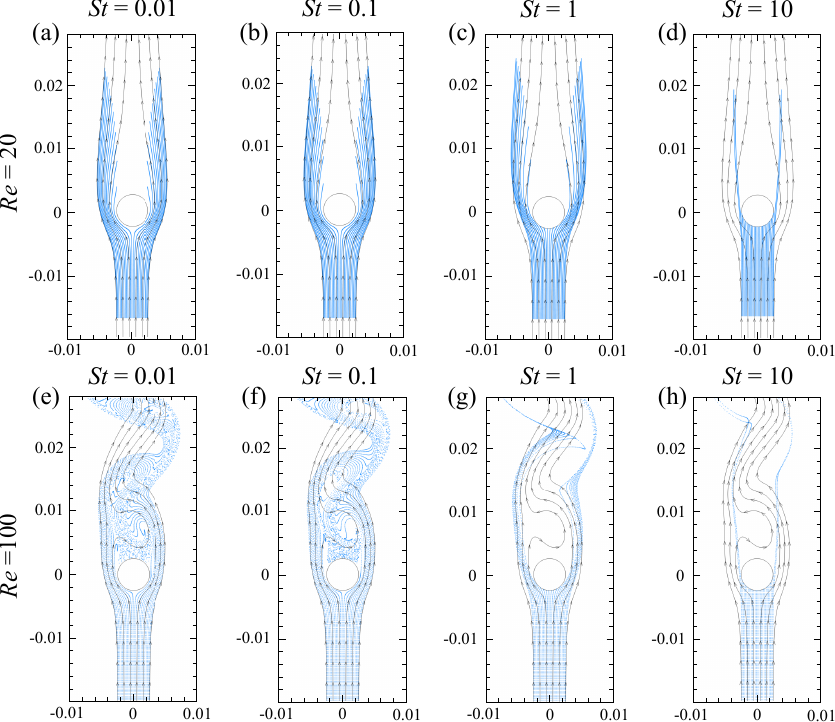}
    \caption{Snapshots of the particle motion at different operating conditions: (a)-(d) refer to those at $Re_\textrm{cyl}=20$,~$St=0.01$,~0.1,~1,~10; (e)-(f) refer to those at $Re_\textrm{cyl}=100$,~$St=0.01$,~0.1,~1,~10. Particles are highlighted by blue dots while the gray lines with arrows represent the streamlines.}
    \label{fig:tracked particle visualization}
\end{figure*}

Fig.~\ref{fig:tracked particle visualization} illustrates particle trajectories under various operating conditions. In general, particles with small diameters-corresponding to small Stokes number ($St=0.01$,~0.1)-follow the streamlines closely. In contrast, particles with large Stokes number ($St=1$,~10) deviate significantly from the flow, showing diminished responsiveness to fluid motion, which is consistent with the inertial impaction theory. Beyond the Stokes number, the Reynolds number also plays a crucial role. As the Reynolds number increases, a Kármán vortex street develops in the downstream of the domain. These vortices influence particle motion in two distinct way. On one hand, the enhanced flow mixing promotes particle accumulation on the downstream side of the cylinder, as shown in Fig.~\ref{fig:tracked particle visualization}(e) and (f). On the other hand, the centrifugal forces associated with these eddies can prevent particles from entering the vortex cores, leading to a blocking effect illustrated in Fig.~\ref{fig:tracked particle visualization}(g).

\begin{figure*}
    \centering
    \includegraphics[width=0.7\linewidth]{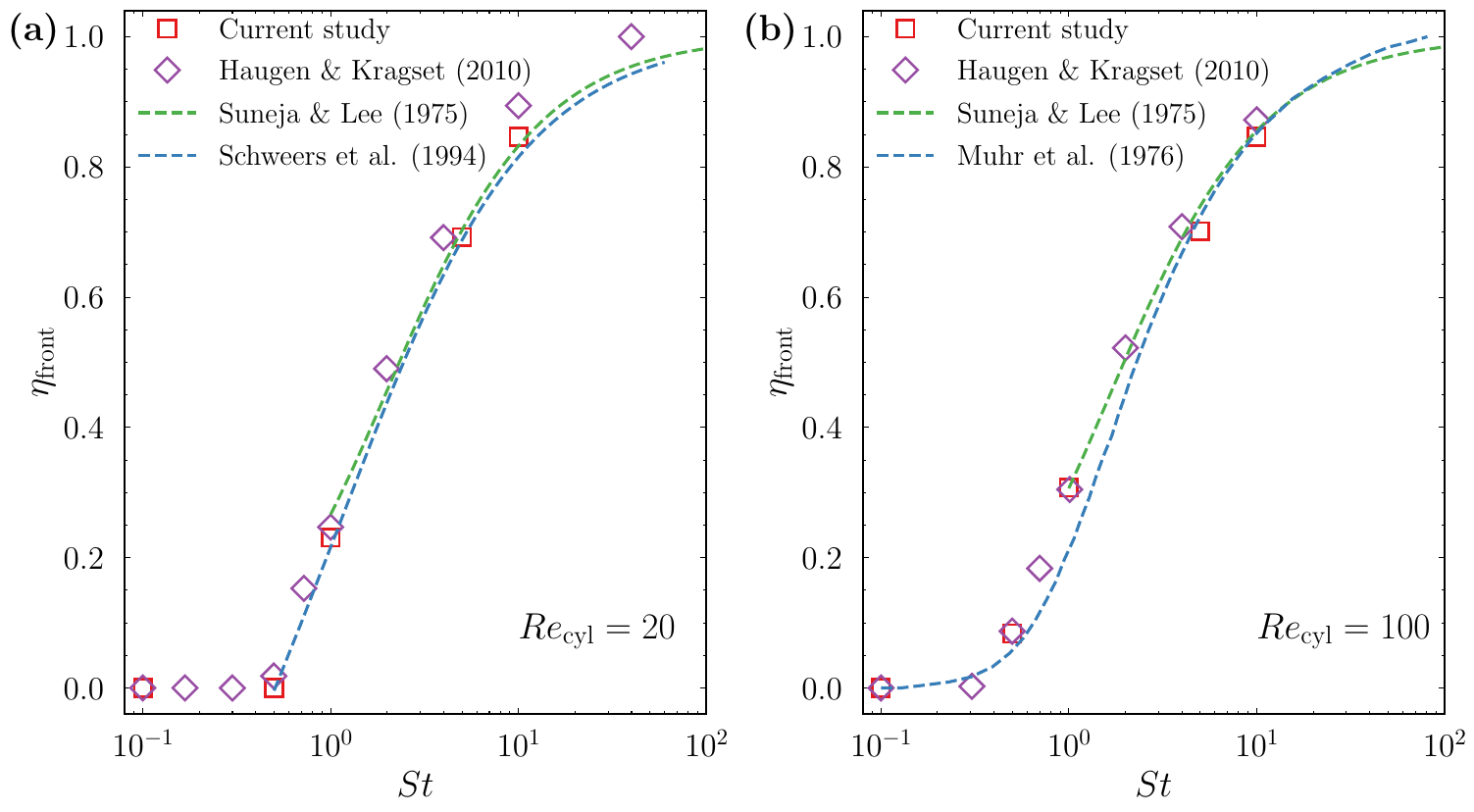}
    \caption{Variation of front-side impaction efficiency with Stokes number under different Reynolds numbers: $Re_\textrm{cyl}=20$ for (a) and $Re_\textrm{cyl}=100$ for (b). Simulation results are compared with previous studies \cite{haugen2010particle}. Reproduced with permission from Journal of Fluid Mechanics 661, 239–261 (2010). Copyright Cambridge University Press 2010}
    \label{fig:front impact efficiency}
\end{figure*}

Based on the particle–collision information obtained from the UDF output, we establish the relationship between the front-side impaction efficiency and the Stokes number under different Reynolds numbers, as shown in Fig.~\ref{fig:front impact efficiency}. We then compare these results with previous studies \cite{haugen2010particle,suneja1974aerosol,schweers1994experimental,muhr1976theoretical} based on theoretical derivations, numerical calculations, and experimental observations, and the agreement is found to be satisfactory.

\section{Results and discussion}
\label{sec:results and discussion}

\subsection{Discretization error analysis and model validation}
\label{sec:discretization error analysis and model validation}
Before conducting the subsequent simulations, both discretization error analysis and model validation were performed to ensure the accuracy and reliability of the numerical model. To simplify the problem, we adopted a method called the cold flow test, inspired by similar practices on actual engine test bench, to focus on observing flow behavior and pressure drop. The working fluid is air with a density of $1.225~\mathrm{kg.m^{-3}}$ and a viscosity of $1.79\times10^{-5}~\mathrm{Pa.s}$. The inlet of the computational domain is set to be a velocity one with $u = 0.0183~\mathrm{m.s^{-1}}$, which leads to $Re_H = 1$. Meanwhile, the outlet is defined as a pressure one with a zero static pressure. The top and bottom sides of the domain are set to be symmetric walls, while all surfaces of porous medium is defined as no-slip and no-penetration walls. Three meshes, from coarse to fine, with different representative grid sizes $\Delta h$ are selected as the objects of current error analysis. The pressure drop between the inlet and outlet of the domain is chosen as the physical quantity to be evaluated. To quantify the discretization errors introduced by different grid resolutions, we adopted the method pioneered by Roache \cite{roache1997quantification}, in which a grid convergence index (GCI) is introduced as a criterion for assessing uncertainty. Table~\ref{tab:gci} summaries the results of current analysis. The GCI value for the last refinement ($\mathrm{GCI}_{12}$) indicates that the uncertainty in the fine-grid solution due to spatial discretization is $0.86\%$, suggesting that discretization error is minor and further refinement is not needed. Therefore, considering the limit of computational resources, medium-refined mesh with $\num{659073}$ grids is selected in the following simulations.
\begin{table}[ht]
    \caption{Discretization error analysis using GCI.}
    \label{tab:gci}
    \begin{ruledtabular}
    \begin{tabular}{lll}
        Parameter & Value & Unit\\
        \hline
        $N_1,~N_2,~N_3$ & $\num{1020733};~\num{659073};~\num{441898}$ & - \\
        $r_{12}$ & $1.25$ & - \\
        $r_{23}$ & $1.22$ & - \\
        $\varphi_{1}$ & $22.59$ & $\mathrm{Pa}$ \\
        $\varphi_{2}$ & $22.84$ & $\mathrm{Pa}$ \\
        $\varphi_{3}$ & $22.92$ & $\mathrm{Pa}$ \\
        $p$ & $4.21$ & - \\
        $\varphi_\textrm{ext}$ & $22.75$ & $\mathrm{Pa}$ \\
        $\mathrm{GCI}_{12}$ & $0.86\%$ & - \\
    \end{tabular}
    \end{ruledtabular}
\end{table}

Due to the inherent complexity and unique characteristic of this porous medium, experimental acquisition of its pressure data is not feasible. To address this challenge, an equivalent representative elementary volume (REV) model is constructed to validate current pore-scale model. The effectiveness of this approach has been validated \cite{tang2020connection}, particularly in cases where the pore structure is an artificial or otherwise idealized model. The REV-scale model can be treated as a macroscopic one, which ignores the detailed micro-structure of porous medium and focuses on its effective macroscopic behavior through averaged physical properties. In our case, an additional source term Eq.~(\ref{eq:Darcy's law}) is then incorporated into Eq.~(\ref{eq:N-S eqn}) to achieve this,
\begin{equation}
    \boldsymbol{S} = -\frac{\mu_\textrm{f}}{K} \boldsymbol{u}, \label{eq:Darcy's law}
\end{equation}
where $K$ is the permeability of the porous media, which is assumed to be isotropic and can be estimated by the Kozeny-Carman equation Eq.~(\ref{eq:Kozeny-Carman}),
\begin{equation}
    K = \frac{\phi^3}{(1 - \phi)^2} \frac{d_{\textrm{eff}}^2}{180}, \label{eq:Kozeny-Carman}
\end{equation}
where $\phi$ is the porosity of the porous media and $d_{\textrm{eff}}$ is the effective grain size.

\begin{figure}[ht]
    \centering
    \includegraphics[width=1.0\linewidth]{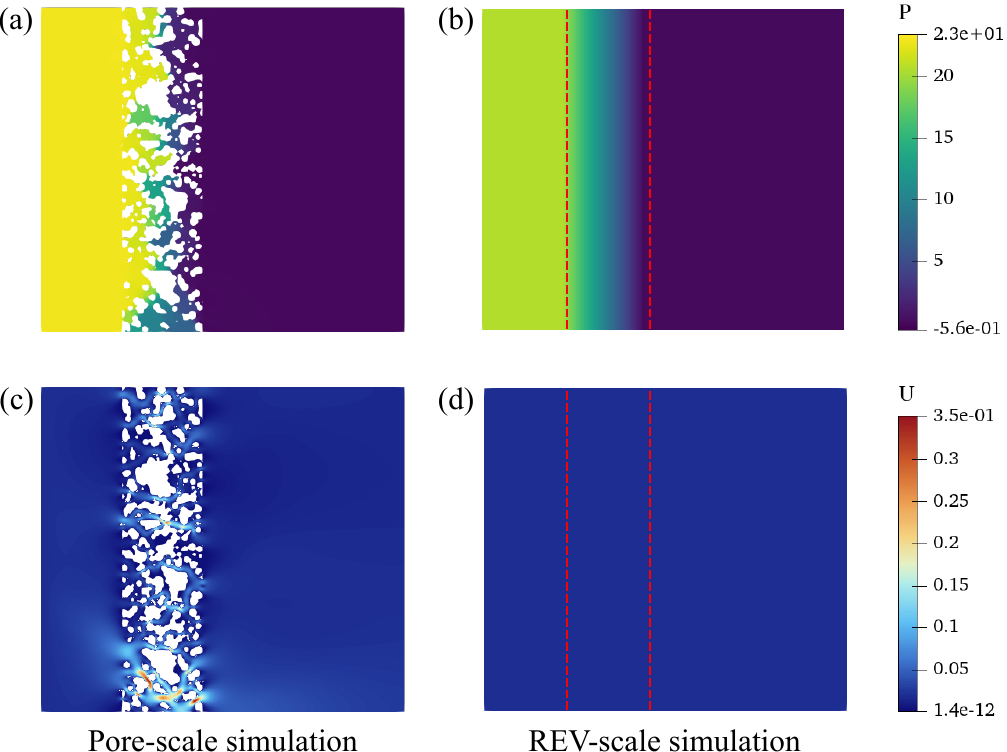}
    \caption{Comparisons of the pressure (a, b)  and velocity (c, d) distribution obtained from pore-scale (a, c) and REV-scale (b, d) simulation. Red dashed lines in (b) and (d) indicate the boundaries of porous zone.}
    \label{fig:pore-scale and REV-scale}
\end{figure}

A quantitative comparison of pressure and velocity distribution from pore-scale and REV-scale approaches is described in Fig.~\ref{fig:pore-scale and REV-scale}. The two results differ in absolute magnitudes but show a consistent overall trend. The differences can be attributed to the assumptions adopted in the REV-scale model. Hence, the process of model validation is completed.

\subsection{Temperature-dependent regeneration process}
\label{sec:temperature-dependent regeneration process}
In this section, we focus on the transport phenomena involved in the simulated regeneration process at an inlet temperature of $600~\mathrm{K}$. This temperature is chosen because it characterizes the exhaust temperature under the regular operation of a diesel engine and also marks the onset of the CDPF passive regeneration process \cite{zhang2020study}, providing a baseline for evaluating our model's capability to capture the details of soot oxidation during this stage.

\begin{figure*}
    \centering
    \includegraphics[width=0.85\linewidth]{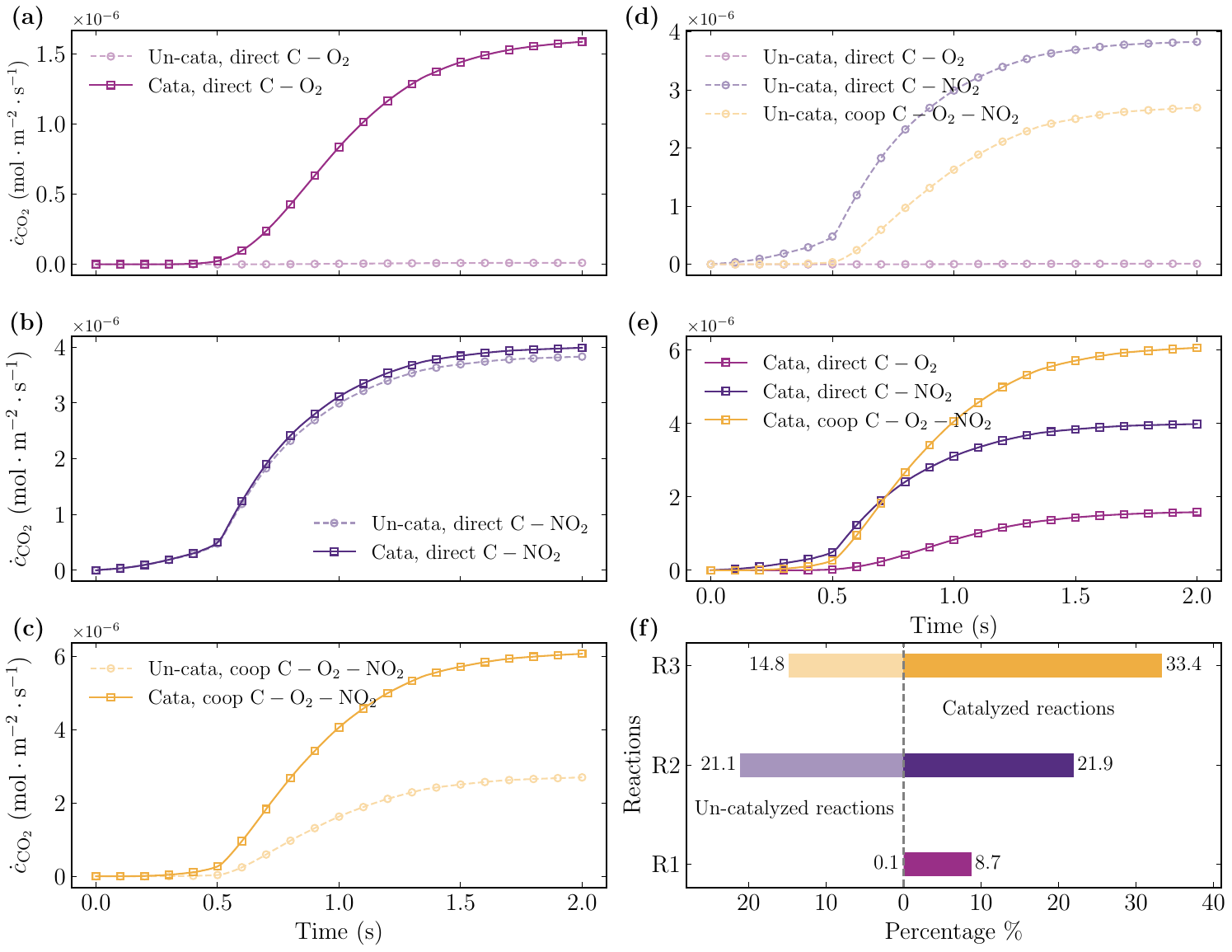}
    \caption{Time-evolved molar production rates of $\textrm{CO}_2$ and contribution to soot regeneration for different heterogeneous chemical reactions in CDPF at $600~\mathrm{K}$: (a)-(c): Effect of catalyst on molar production rates of $\textrm{CO}_2$ for each heterogeneous reaction; (d): Effect of reactants on molar production rates of $\textrm{CO}_2$ for un-catalyzed heterogeneous reactions; (e): Effect of reactants on molar production rates of $\textrm{CO}_2$ for catalyzed heterogeneous reactions; (f): Comparison of percent contribution to regeneration between six heterogeneous reactions at $t=2~\mathrm{s}$.}
    \label{fig:prod rate co2 600K}
\end{figure*}

\begin{figure*}
    \centering
    \includegraphics[width=0.85\linewidth]{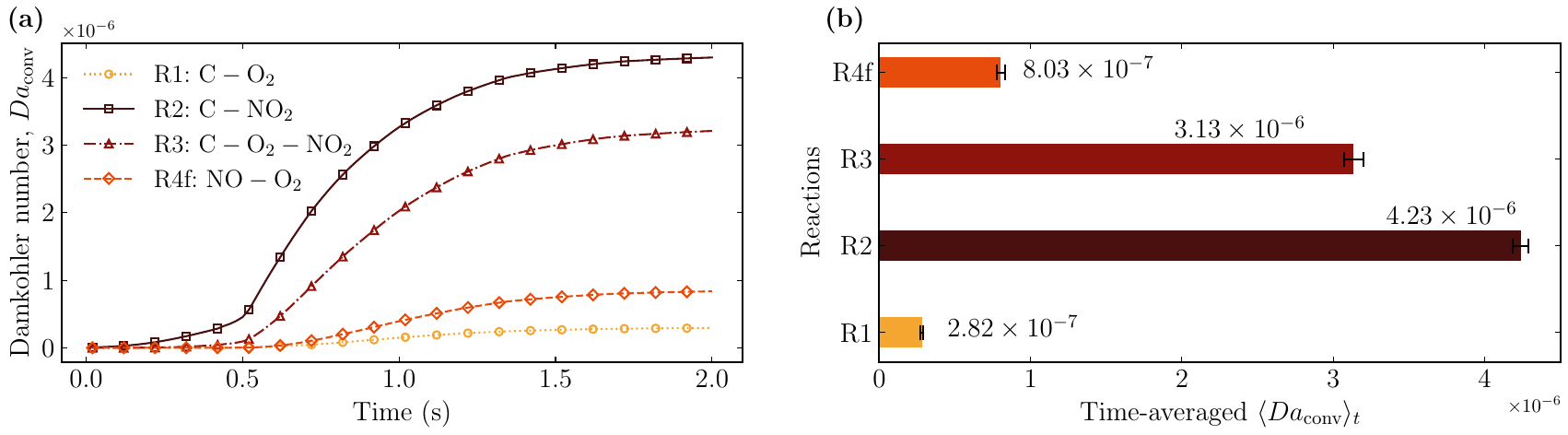}
    \caption{Comparison of the convection Damköhler numbers $Da_\textrm{conv}$ for all chemical reactions within CDPF: (a) Temporal distribution of $Da_\textrm{conv}$ for 4 reactions; (b) Time-averaged $\langle Da_\textrm{conv} \rangle_t$ for 4 reactions from $t=1~\mathrm{s}$ to $t=2~\mathrm{s}$. Both the catalyzed and un-catalyzed reactions are considered in each heterogeneous reactions, while only forward reaction is taken into account in the homogeneous reaction.}
    \label{fig:damkohler number 600k}
\end{figure*}

\begin{figure}
    \centering
    \includegraphics[width=0.85\linewidth]{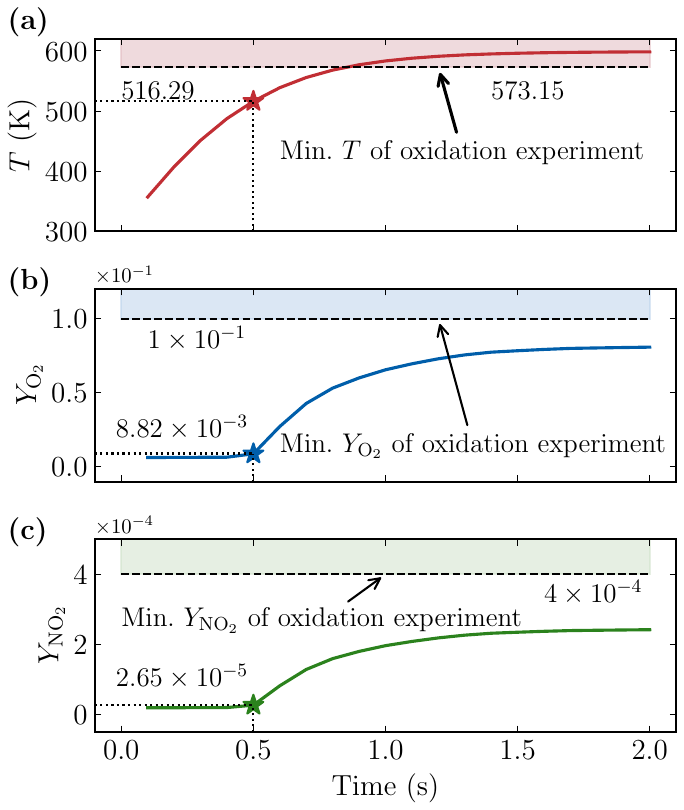}
    \caption{Time-evolved area-weighted averaged temperature (a) and mass fraction of $\textrm{O}_2$ (b) and $\textrm{NO}_2$ (c) in CDPF porous zone. The black dashed line in each subplot indicates the lower limit of corresponding physical quantity employed in a previous experimental study \cite{jeguirim2009kinetics}. The value of each physical quantity at $t=0.5~\mathrm{s}$ is highlighted by a star marker. Reproduced with permission from Journal of Chemical Technology \& Biotechnology: International Research in Process, Environmental \& Clean Technology 84, 770–776 (2009). Copyright 2009 Society of Chemical Industry.}
    \label{fig:area-weighted averaged physical quantity 600K}
\end{figure}

Fig.~\ref{fig:prod rate co2 600K} illustrates the evolution of the molar production rates of $\textrm{CO}_2$ with time for six heterogeneous chemical reactions within CDPF at $600~\mathrm{K}$. The effect of catalyst on each reaction is analyzed through Fig.~\ref{fig:prod rate co2 600K}(a)-(c). It is evident that different reactions exhibit markedly different sensitivities to the catalyst. For the direct $\textrm{C}-\textrm{O}_2$ and cooperative $\textrm{C}-\textrm{O}_2-\textrm{NO}_2$ reactions, the presence of catalyst accelerates the reaction process greatly. However, for the direct $\textrm{C}-\textrm{NO}_2$ reactions, the catalyst seems to have minor effect on the chemical kinetics. These results are consistent with previous findings from isothermal soot oxidation experiments \cite{jeguirim2009kinetics}. We observe that before $0.5~\mathrm{s}$, little $\textrm{CO}_2$ is produced from any of the reactions. This could be explained by the fact that the reactants had not been fully transported from the upstream into the porous media and the temperature at the porous zone remains relatively low, as shown in Fig.~\ref{fig:area-weighted averaged physical quantity 600K}. Fig.~\ref{fig:prod rate co2 600K}(d) compares the molar production rates of $\textrm{CO}_2$ in three un-catalyzed heterogeneous reactions. At the same moment, the carbon dioxide produced by the direct $\textrm{C}-\textrm{NO}_2$ reaction is higher than that in the cooperative $\textrm{C}-\textrm{O}_2-\textrm{NO}_2$, approximately 1.5 times at $t=2~\mathrm{s}$, whereas the molar production rate of $\textrm{CO}_2$ remains zero for un-catalyzed direct $\textrm{C}-\textrm{O}_2$ reaction throughout the entire simulation. The reason for the latter is fairly straightforward. In general, the direct $\textrm{C}-\textrm{O}_2$ reaction requires a relatively high exhaust temperature (about 800 to 1000~$\mathrm{K}$) to sustain its rapid reaction rate in the absence of the Pt catalyst. However, under our operating conditions, such a requirement cannot be met, as shown in Fig.~\ref{fig:area-weighted averaged physical quantity 600K}(a). This explains why the $\textrm{CO}_2$ molar production rate curve of un-catalyzed direct $\textrm{C}-\textrm{O}_2$ reaction looks flat in Fig.~\ref{fig:prod rate co2 600K}(a) and (d). In contrast, the result for the former is somewhat unexpected. In previous studies based on oxidation experiments, researchers reported that under the same temperature and without the aid of catalyst, the reaction rate of the cooperative pathway should be fast than that of the direct one \cite{jeguirim2009kinetics}. Nevertheless, the results from our present simulation appear to contradict this conclusion. The underlying reason is that, in the experimental studies used to determine the kinetic parameters of soot oxidation, the two reaction pathways are evaluated separately. In contrast, our simulation allows these pathways to proceed simultaneously, a distinct strength of our model to capture intrinsic coupling between different reactions, thereby introducing competition between them. To quantify such competition, a dimensionless group is introduced, named the convection Damköhler number \cite{rehage2021first}, which is defined as Eq.~(\ref{eq:convection Da}),
\begin{equation}
    Da_\textrm{conv} = \frac{\tau_\textrm{conv}}{\tau_\textrm{chem}}, \label{eq:convection Da}
\end{equation}
where $\tau_\textrm{conv}$ represents the convection time scale, and $\tau_\textrm{chem}$ denotes the chemical reaction time scale \cite{wartha2021characteristic}, and the definitions of them are listed through Eq.~(\ref{eq:convection time scale}) and (\ref{eq:chemical reaction time scale}),
\begin{equation}
    \tau_\textrm{conv} = \frac{H}{u_0}, \label{eq:convection time scale}
\end{equation}
\begin{equation}
    \tau_\textrm{chem} = \frac{c_\textrm{tot}}{\sum_{i=1}^{N_\textrm{R.H.S}} \nu_i v_\textrm{r}}. \label{eq:chemical reaction time scale}
\end{equation}

Fig.~\ref{fig:damkohler number 600k} compares the convection Damköhler numbers for all chemical reactions in CDPF. Clearly, $Da_\textrm{conv}$ of reaction 2 is larger than that of reaction 3, which implies that the reaction rate of cooperative $\textrm{C}-\textrm{O}_2-\textrm{NO}_2$ is actually slower than that of direct $\textrm{C}-\textrm{NO}_2$. Considering that these reactions proceed in parallel and the upstream $\textrm{NO}_2$ concentration is low, as well as the limited amount of $\textrm{NO}_2$ produced through the homogeneous reaction (see in Fig.~\ref{fig:damkohler number 600k}), the direct $\textrm{C}-\textrm{NO}_2$ pathway consistently holds the advantage in this competition. This explains why the results in Fig.~\ref{fig:prod rate co2 600K}(d) diverge from those reported in experimental studies.

Fig.~\ref{fig:prod rate co2 600K}(e) compares the evolution of $\textrm{CO}_2$ molar production rates with time between three catalyzed heterogeneous reactions. In the presence of the catalyst, the $\textrm{CO}_2$ formation via the cooperative pathway is slightly lower than that via the direct pathway before $0.7~\textrm{s}$. However, after this point, the former pathway produces more $\textrm{CO}_2$ than the latter, a trend that has also been observed in experimental studies.

Fig.~\ref{fig:prod rate co2 600K}(f) demonstrates the percent contribution to soot oxidation for different heterogeneous chemical reactions at $t=2~\textrm{s}$. In general, at an inlet temperature of $600~\mathrm{K}$, the catalyst significantly influences soot oxidation. Catalyzed reactions contribute $64\%$ to the regeneration, while un-catalyzed reactions account for only $36\%$. $\textrm{NO}_2$ plays a dominant role due to its relatively high reaction rate at low temperatures, as shown in Fig.~\ref{fig:damkohler number 600k}. In contrast, $\textrm{O}_2$ has a negligible effect on the overall oxidation process.
\subsection{Particle deposition}
\label{sec:particle deposition}
In this section, we study the particle deposition behavior under the reactive flow conditions described above. A total of \num{20420} particles were injected into the computational domain from the specified cross-section, of which up to \num{19979} were deposited in the porous medium, resulting in a deposition rate of $97\%$. 

\begin{figure*}
    \centering
    \includegraphics[width=0.65\linewidth]{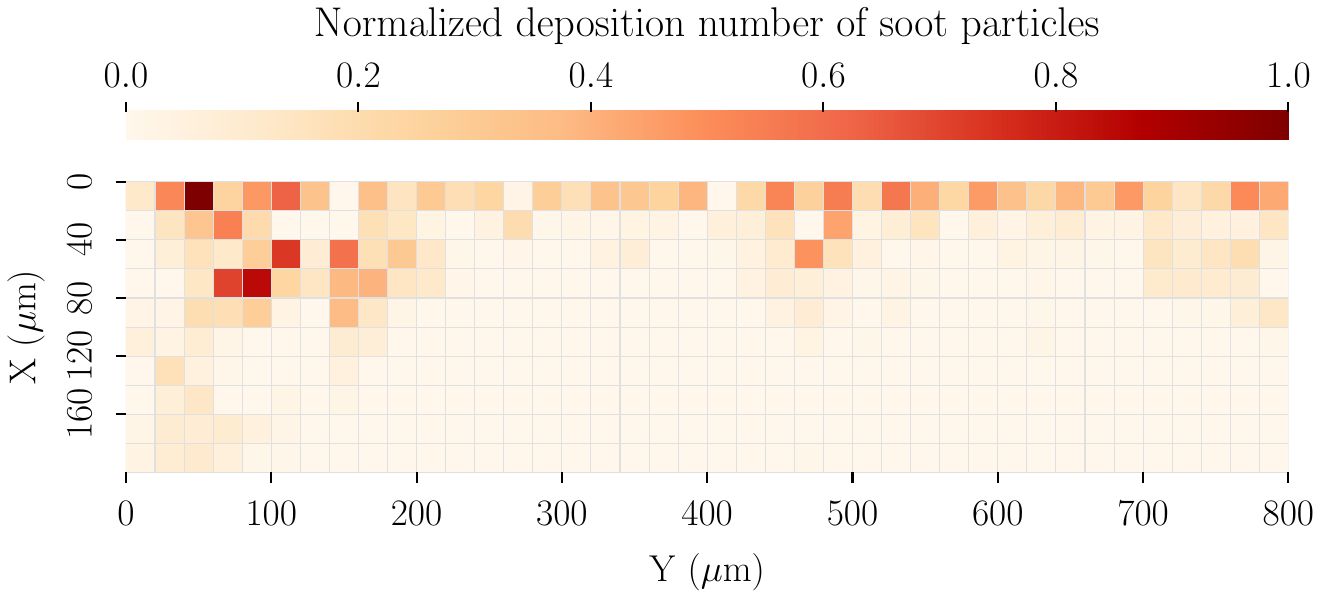}
    \caption{Spatial distribution of particle deposition within CDPF porous media.}
    \label{fig:deposit heatmap}
\end{figure*}

\begin{figure}[ht]
    \centering
    \includegraphics[width=1.0\linewidth]{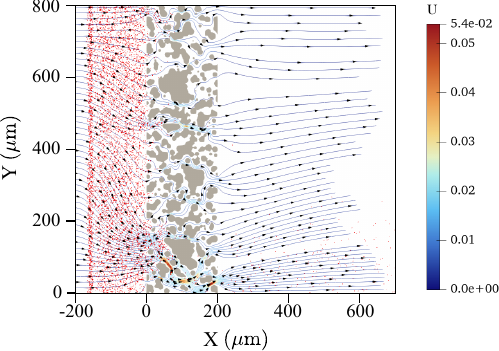}
    \caption{Snapshot of particle trajectories and streamlines at $t=0.8~\mathrm{s}$. Particles are highlighted by red spheres while streamlines are colored by velocity magnitudes. Black arrows on each streamline represent the velocity vectors.}
    \label{fig:snapshot particle trajectory}
\end{figure}

Fig.~\ref{fig:deposit heatmap} illustrates the spatial distribution of soot deposition within the CDPF porous wall. In the streamwise direction, the soot particles are predominantly intercepted in the first quarter of the porous region. This sharp concentration gradient is attributed to the high initial filtration efficiency of the wall, resulting in a classic depth-filtration profile consistent with previous findings \cite{huang2024study}. Perpendicular to the flow, significant heterogeneity is observed, particularly the hotspot in the lower region (from 0 to 200~$\mathrm{\mu m}$ in $y$-axis). To explain this result, Fig.~\ref{fig:snapshot particle trajectory} correlates this pattern with the local flow field and particle trajectories. The stochastic geometry induces alternating expansion-contraction channels, which generate steep local velocity gradients. These preferential flow paths facilitate a higher convective particle flux, leading to increased particle throughput. While the deposition probability per particle contact remains consistent with the local filtration mechanism, the increased total throughput of particles leads to the pronounced localized accumulation seen in Fig.~\ref{fig:deposit heatmap}.

\begin{figure}[ht]
    \centering
    \includegraphics[width=0.8\linewidth]{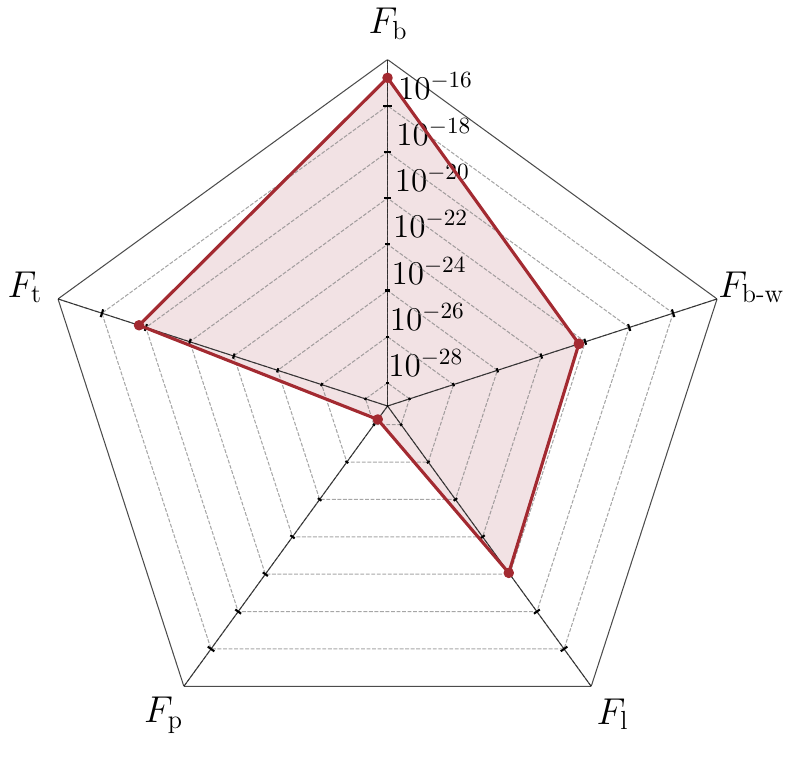}
    \caption{Scale analysis of five forces acting on a soot particle with a diameter of $50~\mathrm{nm}$.}
    \label{fig:scale analysis}
\end{figure}

An interesting observation from our simulations is that nearly all particles that collide with the wall ultimately adhere to the substrate surface. Therefore, the following analysis focuses on the effects of the different impaction mechanisms. To quantify these mechanisms, a scale analysis of several forces acting on a soot particle is conducted. It is worth noting that the drag and virtual mass forces are not considered in the present analysis, as they are in the same direction as fluid flows \cite{wang2024numerical}. The result is illustrated in Fig.~\ref{fig:scale analysis}. Clearly, the Brownian force stands out in particle dynamics, reaching a magnitude of $10^{-15}~\mathrm{N}$. This outcome validates previous theoretical studies on diesel particulate capture mechanisms \cite{konstandopoulos1989wall}, highlighting Brownian diffusion as a key filtration mode. Due to the temperature fluctuation during the oxidation process, the thermophoretic force also plays an important role in the impaction mechanism, with a magnitude of $10^{-18}~\mathrm{N}$. The effects of buoyancy and Saffman's lift on the particle are relatively small, while the influence of the pressure gradient force is almost negligible, which can be explained by the extreme small value of fluid-particle density ratio.

\section{Conclusion}
\label{sec:conclusion}
In this work, we first proposed a workflow for the parameter-driven reconstruction of the CDPF porous structures. Based on this method, a two-dimensional geometry represented of the in-wall CDPF porous media was generated. The geometric non-uniformity of the CDPF porous structure is effectively represented. 

Next, a multiphysics mathematical model which is capable of capturing real-time transport phenomena within the CDPF was established. For the complex chemical reactions within the CDPF, the reaction kinetics and thermochemistry are incorporated into the fundamental governing equations through appropriate boundary conditions or source terms, depending on the reaction types. For the deposition process of soot particles in CDPF, a particle-wall interaction model was developed based on the B-H theory, which interprets the impact-adhere-rebound process physically.

To numerically solve the proposed mathematical model using Fluent, a series of UDFs and UDMs were developed. Several classical benchmarks were employed for code verification. The results show that the established model can accurately capture mass transfer under heterogeneous chemical reactions as well as the particle-wall interaction.

Finally, the proposed numerical model was employed to simulate the regeneration and particulate filtration process of the in-wall CDPF under a normal operating condition. The results provide quantitative insights that distinguish this pore-scale approach from traditional macroscopic models. The model realistically reflects the competitive and synergistic interactions among various chemical pathways. Specifically, it is quantified that the presence of the catalyst accelerates the reaction rate of the direct $\text{O}_2$ pathway by a factor of 87 under low-temperature regeneration conditions. This highlighting the pivotal role of $\text{NO}_2$ and catalytic active sites in lowering the regeneration threshold. For ultra-fine soot particles ($50~\mathrm{nm}$), the simulation reveals that Brownian motion and thermophoretic forces are the dominant mechanisms dictating deposition efficiency. The high sensitivity of these forces to local thermal gradients, captured through the integrated heat and mass transfer framework, underscores the necessity of accounting for non-isothermal effects at the pore scale.

Despite the insights gained from this study, certain limitations remain that define the scope for future investigations. Firstly, the catalytic soot oxidation is currently represented by lumped kinetic parameters rather than explicit modeling of catalytic active sites. Future research will aim to incorporate molecular dynamics simulation to accurately resolve the interactions between soot, gas-phase oxidants, and specific catalyst compositions at the molecular level. Secondly, the monodisperse particle assumption will be replaced by typical trimodal particle size distribution in our future works. In addition, links between the proposed pore-scale model and macroscopic models will be established through engine bench tests, which is of significant importance for guiding CDPF development.


\begin{acknowledgments}
This study was supported by the National Natural Science Foundation of China (No.~52206167 and No.~52472414). We also acknowledge Prof.~Petr~Kočí and Dr.~Hannah~Menke for making their data publicly available.
\end{acknowledgments}

\section*{Data Availability Statement}
The codes and raw post-processing data supporting the findings of this research are publicly available at \href{https://github.com/zhangyujing2001}{GitHub}. Feel free to submit a pull request (PR) on GitHub if needed. Additional data, including the detailed case settings and post-processing UDFs and Python codes, are available from the corresponding author upon reasonable request.

\appendix

\section{Derivation of the rate laws for chemical reactions in CDPF} 
\label{app:rate laws derivation}

In the following part, the general form of the rate law is first introduced, based on which the rate laws for different reaction types within the CDPF are derived.

Eq.~(\ref{eq:general form of rate law}) depicts the general form of the rate law for a chemical reaction,
\begin{equation}
    v_\textrm{r} = k_\textrm{r} \prod_{i=1}^{N_\textrm{L.H.S}}(c_i)^{\alpha_i},
    \label{eq:general form of rate law}
\end{equation}
where $v_\textrm{r}$ is the reaction rate, $c_i$ represents the molar concentration of the $i$-th reactant, $\alpha_i$ stands for the reaction order of the $i$-th reactant, and $k_\textrm{r}$ is the rate constant, which can be determined by the Arrhenius equation Eq.~(\ref{eq:Arrhenius equation}),
\begin{equation}
    \mathrm{ln} k_\textrm{r} = \mathrm{ln} A - \frac{E_a}{R T},
    \label{eq:Arrhenius equation}
\end{equation}
where $A$ is the frequency factor, also known as the pre-exponential factor, $E_a$ is the activation energy, and $R$ is the gas constant.

For a heterogeneous reaction, the concentration of a pure solid is not considered because it remains constant and cannot be varied experimentally \cite{soustelle2011introduction}. Therefore, the rate law for the $j$-th heterogeneous reaction can be expressed as Eq.~(\ref{eq:rate law for heterogeneous react}),
\begin{equation}
    v_{\textrm{r}, j} = A_j^*~\mathrm{exp}\left( -\frac{E_{a,j}}{R T} \right) \prod_{k=1}^{N_\textrm{L.H.S}} (Y_k)^{\alpha_{k,j}} c_{\textrm{tot}},
    \label{eq:rate law for heterogeneous react}
\end{equation}
where $A^*$ is the modified frequency factor ($A^* = \mathrm{SSA}^{-1}A$, where $\mathrm{SSA}$ is the specific surface area of the porous media), $Y$ is the mass fraction of the gas phase reactant, and $c_\textrm{tot}$ is the molar concentration of the mixture, which can be computed by the ideal gas law.

For a reversible homogeneous reaction like R4, the rate law is a little bit complicated, as both the forward and reverse reaction kinetics should be considered. To solve this, a thermodynamic equilibrium constant is introduced and its definition has been given in Eq.~(\ref{eq:thermodynamic equilibrium constant}),
\begin{equation}
    K_\textrm{eq} = \left[ \prod_{i=1}^{N} \left(\frac{p_i}{p^{\plimsoll}} \right)^{\nu_i} \right]_{\textrm{eq}},
    \label{eq:thermodynamic equilibrium constant}
\end{equation}
where $p_i$ is the partial pressure of species $i$ ($p_i = X_i p$, where $X_i$ is the mole fraction of the species $i$ and $p$ is the total pressure), and $p^{\plimsoll}$ is the standard atmosphere. Combining Eqs.~(\ref{eq:general form of rate law}) and (\ref{eq:thermodynamic equilibrium constant}), a rate law for a homogeneous and reversible chemical reaction can be expressed as Eq.~(\ref{eq:rate law for reversible reaction}),
\begin{equation}
    v_\textrm{r} = k_{\textrm{r,forward}} \left( \prod_{i=1}^{N_{\textrm{L.H.S}}} (Y_i)^{\alpha_i} - K_\textrm{eq}^{-1} \prod_{i=1}^{N_{\textrm{R.H.S}}} (Y_i)^{\alpha_i} \right) c_\textrm{tot},
    \label{eq:rate law for reversible reaction}
\end{equation}
where $k_{\textrm{r,forward}}$ is the rate constant of the forward reaction.

The catalyst used in this work is a typical platinum-based commercial catalyst $\textrm{Pt-Al}_2\textrm{O}_3$ \cite{tan2019modeling}, which is widely used in CDPF to promote $\textrm{NO-NO}_2$ conversion and soot oxidation \cite{guan2024experimental}. Table.~\ref{tab:chemical parameters} summarizes the parameters used in the CDPF chemical kinetics and thermochemistry model.
\begin{table*}[ht]
    \caption{Parameters used in the CDPF chemical kinetics and thermochemistry model \cite{tan2019modeling,jeguirim2009kinetics}. Reproduced with permissions from Chemical Engineering Journal 375, 122110 (2019). 2019 Elsevier B.V. All rights reserved; Journal of Chemical Technology \& Biotechnology: International Research in Process, Environmental \& Clean Technology 84, 770–776 (2009). Copyright 2009 Society of Chemical Industry.}
    \label{tab:chemical parameters}
    \begin{ruledtabular}
    \begin{tabular}{>{\raggedright\arraybackslash}p{0.03\linewidth} >{\raggedright\arraybackslash}p{0.3\linewidth} >{\raggedright\arraybackslash}p{0.14\linewidth} >{\raggedright\arraybackslash}p{0.16\linewidth} >{\raggedright\arraybackslash}p{0.16\linewidth}}
        No. & Thermochemical equation ($\Delta_\textrm{r}H^{\plimsoll},~\mathrm{kJ.mol^{-1}}$) & Reaction order ($\alpha_i$,~-) & Frequency factor ($A,~\mathrm{s^{-1}}$) & Activation energy ($E_\textrm{a},~\mathrm{J.mol^{-1}}$) \\
        \hline
        R1 & $\textrm{C(s)} + \textrm{O}_2(\textrm{g}) \rightarrow \textrm{CO}_2(\textrm{g})$,~$(-393.5)$& $\alpha_{\textrm{O}_2} = 0.9$ & $8.50\times10^{7}$ & $1.64\times10^{5}$ \\
          & If \textbf{catalyzed}, then, & $\alpha_{\textrm{O}_2} = 0.3$ & $1.19\times10^{5}$ & $1.14\times10^{5}$ \\
        R2 & $\textrm{C(s)} + 2\textrm{NO}_2(\textrm{g}) \rightarrow \textrm{CO}_2(\textrm{g})+2\textrm{NO}(\textrm{g})$,~$(-279.3)$ & $\alpha_{\textrm{NO}_2} = 0.6$ & $0.48$ & $2.67\times10^{4}$ \\
          & If \textbf{catalyzed}, then, & $\alpha_{\textrm{NO}_2} = 0.6$ & $0.51$ & $2.68\times10^{4}$ \\
        R3 & $\textrm{C(s)} + \textrm{NO}_2(\textrm{g}) + 0.5\textrm{O}_2(\textrm{g}) \rightarrow \textrm{CO}_2(\textrm{g})+\textrm{NO}(\textrm{g})$,~$(-336.4)$ & $\alpha_{\textrm{O}_2} = 0.3,\newline~\alpha_{\textrm{NO}_2} = 1.0$ & $1.395\times10^{3}$ & $4.78\times10^{4}$ \\
          & If \textbf{catalyzed}, then, & $\alpha_{\textrm{O}_2} = 0.3,\newline~\alpha_{\textrm{NO}_2} = 0.4$ & $51.4$ & $5.22\times10^{4}$ \\
        R4 & $\textrm{NO}(\textrm{g}) + 0.5\textrm{O}_2(\textrm{g}) \rightleftharpoons \textrm{NO}_2(\textrm{g})$,~$(-57.1)$ (Catalyzed reaction)& $\alpha_{\textrm{O}_2} = 0.5$,\newline~$\alpha_{\textrm{NO}} = 1.0$,\newline~$\alpha_{\textrm{NO}_2} = 1.0$ & $3.61\times10^{6}$ & $8.18\times10^{4}$ \\
    \end{tabular}
    \end{ruledtabular}
\end{table*}

\section{Forces acting on a soot particle}
\label{app:forces acting on a soot particle}
The definition of each force in Eq.~(\ref{eq:particle motion eqn}) is listed as follows.

Considering that the research subject is the ultra-fine soot particle, we apply the Stokes-Cunningham drag model, which is expressed in Eq.~(\ref{eq:stokes-cunningham drag model}),
\begin{equation}
    \boldsymbol{F}_\textrm{d} = m_\textrm{p} \frac{\rho_\textrm{p} d_\textrm{p}^2 C_c}{18 \mu_\textrm{f}} (\boldsymbol{u} - \boldsymbol{u}_p),
    \label{eq:stokes-cunningham drag model}
\end{equation}
where $\rho_\textrm{p}$ is the particle density, $d_\textrm{p}$ is the particle diameter, and $C_c$ is the Cunningham correction factor, expressed by Eq.~(\ref{eq:Cunningham correction factor}),
\begin{equation}
    C_c = 1 + Kn_\textrm{p} \left(1.257 + 0.4e^{-(1.1 / Kn_\textrm{p})}\right),
    \label{eq:Cunningham correction factor}
\end{equation}
where $Kn_\textrm{p} = 2 \lambda / d_\textrm{p}$ is the particle Knudsen number, and $\lambda$ is the mean free path of the mixture. 

The summation of buoyancy and weight is computed by Eq.~(\ref{eq:buoyancy and weight}),
\begin{equation}
    \boldsymbol{F}_\textrm{b-w} = m_\textrm{p} \left(1 - \frac{\rho_\textrm{f}}{\rho_\textrm{p}}\right) \boldsymbol{g}.
    \label{eq:buoyancy and weight}
\end{equation}

Next is the Brownian force $\boldsymbol{F}_b$, the magnitude is defined as Eq.~(\ref{eq:Browian force}),
\begin{equation}
    F_{\textrm{b}_i} = m_\textrm{p} \zeta_i \sqrt{\frac{\pi S_0}{\Delta t}},
    \label{eq:Browian force}
\end{equation}
where $\zeta_i$ represents a standard Gaussian white noise process, and $S_0$ denotes the spectral intensity, expressed as Eq.~(\ref{eq:spectral intensity}),
\begin{equation}
    S_0 = \frac{216 \nu_f k_B T}{\pi^2 \rho_\textrm{f} d_\textrm{p}^5 \left(\rho_\textrm{f}/\rho_\textrm{p}\right)^2 C_c},
    \label{eq:spectral intensity}
\end{equation}
where $k_B$ is the Boltzmann constant.

Due to the temperature gradient of the fluid phase, the thermophoretic force $\boldsymbol{F}_\textrm{t}$ is considered, which is computed through Eq.~(\ref{eq:thermophoretic force}),
\begin{equation}
    \boldsymbol{F}_\textrm{t} = - D_T \frac{\nabla T}{T},
    \label{eq:thermophoretic force}
\end{equation}
where $D_T$ is the thermophoretic coefficient and its value can be obtained through Talbot formula Eq.~(\ref{eq:Talbot formula}),
\begin{equation}
    D_T = \frac{6 \pi d_\textrm{p} \mu_\textrm{f}^2 C_s (k_\textrm{f}/k_\textrm{p} + C_t Kn_\textrm{p})}{\rho_\textrm{f} (1 + 3 C_m Kn_\textrm{p})(1 + 2k_\textrm{f}/k_\textrm{p} + 2C_t Kn_\textrm{p})},
    \label{eq:Talbot formula}
\end{equation}
where $C_m = 1.14$,~$C_s = 1.17$,~$C_t = 2.18$.

The virtual mass force can be written as Eq.~(\ref{eq:virtual mass}),
\begin{equation}
    \boldsymbol{F}_\textrm{v} = m_\textrm{p} \frac{\rho_\textrm{f}}{\rho_\textrm{p}}\left(\boldsymbol{u}_\textrm{p} \nabla \boldsymbol{u} - \frac{d \boldsymbol{u}_\textrm{p}}{d t} \right) C_{\textrm{vm}},
    \label{eq:virtual mass}
\end{equation}
where $C_{\textrm{vm}}$ is the virtual mass coefficient with a value of $0.5$.

The pressure gradient force $\boldsymbol{F}_\textrm{p}$ is also included in our model, which is expressed as Eq.~(\ref{eq:pressure gradient force}),
\begin{equation}
    \boldsymbol{F}_\textrm{p} = m_\textrm{p} \frac{\rho_\textrm{f}}{\rho_\textrm{p}} \boldsymbol{u} \nabla \boldsymbol{u}.
    \label{eq:pressure gradient force}
\end{equation}

The last one is the Saffman's lift force $\boldsymbol{F}_\textrm{l}$, obtained by Eq.~(\ref{eq:Saffman's lift}),
\begin{equation}
    \boldsymbol{F}_\textrm{l} = m_\textrm{p} \frac{5.098 \nu_\textrm{f}^{0.5} \rho_\textrm{f} d_{ij}} {\rho_\textrm{p} d_\textrm{p} (d_{lk} d_{kl})^{0.25}} (\boldsymbol{u} - \boldsymbol{u}_\textrm{p}),
    \label{eq:Saffman's lift}
\end{equation}
where $d_{ij}$ is the deformation tensor.

\section{User-defined functions and user-defined memories}
\label{app:udfs and udms}
Table.~\ref{tab:udfs and udms} shows the UDFs and UDMs used in the numerical model. Note that all post-processing UDFs are not concluded in this list.

\begin{table*}
    \caption{Summary of user-defined functions and user-defined memories in numerical models.}
    \label{tab:udfs and udms}
    \begin{ruledtabular}
    \begin{tabular}{ lll }
        Type & Name & Feature \\
        \hline
        $\texttt{DEFINE\_PROFILE}$ & $\texttt{massFluxOxygen}$ & A Neumann B.C. for computing mass flux of $\textrm{O}_2$ at wall boundaries \\
         & $\texttt{massFluxNitrogenDioxide}$ & A Neumann B.C. for computing mass flux of $\textrm{NO}_2$ at wall boundaries \\
         & $\texttt{massFluxNitricOxide}$ & A Neumann B.C. for computing mass flux of $\textrm{NO}$ at wall boundaries \\
         & $\texttt{massFluxCarbonDioxide}$ & A Neumann B.C. for computing mass flux of $\textrm{CO}_2$ at wall boundaries \\
         & $\texttt{heatFlux}$ & A Neumann B.C for computing heat flux at wall boundaries \\
        $\texttt{DEFINE\_SOURCE}$ & $\texttt{oxygenSource}$ & Volumetric source term in the species transport equation (Zone 2) for $\textrm{CO}_2$ \\
         & $\texttt{nitrogenDioxideSource}$ & Volumetric source term in the species transport equation (Zone 2) for $\textrm{NO}_2$ \\
         & $\texttt{nitricOxideSource}$ & Volumetric source term in the species transport equation (Zone 2) for $\textrm{NO}$ \\
         & $\texttt{heatReleaseFromR4}$ & Volumetric source term in the energy equation (Zone 2) \\
        $\texttt{DEFINE\_ON\_DEMAND}$ & $\texttt{initUDMs}$ & Initialize UDMs for DPM model \\
        $\texttt{DEFINE\_DPM\_BC}$ & $\texttt{impactDepositRebound}$ & A boundary condition for particle-wall interactions \\
        $\texttt{UDMs}$ & $\texttt{UDM\_IMPACT}$ & Record the number of impaction \\
         & $\texttt{UDM\_DEPOSIT}$ & Record the number of deposition \\
    \end{tabular}
    \end{ruledtabular}
\end{table*}

\bibliography{aipmain}

\end{document}